\documentclass[final,3p,times]{elsarticle}

\usepackage{amssymb}
\usepackage{amsmath}
\usepackage{float}
\usepackage{multirow}
\usepackage{algorithm}
\usepackage{algpseudocode}
\usepackage{gensymb}
\usepackage[hidelinks]{hyperref}
\usepackage{listings}
\usepackage{subcaption}
\usepackage{adjustbox}
\usepackage{bm}
\usepackage{xcolor}

\usepackage{caption}
\usepackage{graphicx}
\usepackage{latexsym}
\usepackage{times}
\usepackage[pagewise]{lineno}

\usepackage{fmtcount}
\usepackage{marvosym}
\usepackage{bm}
\usepackage{chemformula}
\usepackage{adjustbox}
\usepackage{epstopdf}

\usepackage{algorithm}
\usepackage{algpseudocode}
\usepackage{gensymb}
\usepackage[hidelinks]{hyperref}
\usepackage{listings}
\usepackage{adjustbox}
\usepackage{xcolor}
\usepackage{booktabs}

\journal{}

\begin{document}
\sloppy
\begin{frontmatter}



\title{Accuracy and scalability of asynchronous compressible flow solver for transitional flows}

\author[inst1,inst2]{Aswin Kumar Arumugam}
\author[inst1]{Shubham Kumar Goswami}
\author[inst2]{Nagabhushana Rao Vadlamani}
\author[inst1]{Konduri Aditya\corref{cor1}}

\affiliation[inst1]{organization={Department of Computational and Data Sciences, Indian Institute of Science},
            city={Bengaluru},
            postcode={560012},
            state={KA},
            country={India}}

\affiliation[inst2]{organization={Department of Aerospace Engineering, Indian Institute of Technology Madras},
            city={Chennai},
            postcode={600036},
            state={TN},
            country={India}}

\cortext[cor1]{Corresponding author.}

\begin{abstract}
To overcome the communication bottlenecks observed in state-of-the-art parallel time-dependent flow solvers at extreme scales, an asynchronous computing approach that relaxes communication and synchronization at a mathematical level was previously developed. This approach preserves high-order accuracy of computations near processing element boundaries using asynchrony-tolerant (AT) schemes while significantly improving the scalability. The numerical properties of the AT schemes were studied based on simple linear and nonlinear partial differential equations (PDEs) in previous works. Allowing asynchrony in numerical schemes can minimize communication overheads in a parallel setting in two ways: one that avoids communication over a few predetermined time steps, and the other that initiates communications without enforcing synchronization. In this study, the asynchronous algorithms are incorporated into the high-order compressible flow solver COMP-SQUARE, which solves practically relevant flow problems in complex geometries in a multi-block framework. The numerical efficacy and scalability of the two asynchronous algorithms are demonstrated for three test cases: isentropic advection of a vortex, the Taylor-Green vortex, and a much more sensitive case of the flow transitioning on a NACA0012 airfoil. Speed-ups of up to $4\times$ with respect to the baseline synchronous algorithm are observed in the scaling experiments performed on up to 18,432 cores. The results of this study demonstrate the applicability of AT schemes on established CFD solvers to improve scalability at extreme scales as the scientific computing environment moves to the exascale era.
\end{abstract}



\end{frontmatter}


\section{Introduction}

Direct numerical simulations (DNS) of practically relevant flow problems are computationally expensive and, therefore, require massive parallel computing resources. Current state-of-the-art DNS solvers often use high-order finite-difference schemes to approximate spatial derivatives and multistage schemes such as Runge-Kutta or multistep schemes such as Adams-Bashforth for temporal integration \cite{Chen_CSD_2009,Jenkins_DNS_1999,Desjardins_JCP_2008}. While advancing in time, stencil operations are performed at grid points where the spatial derivatives are computed based on data from neighboring points. Implementing such methods in a parallel setting requires domain decomposition with a number of processing elements (PEs). This introduces data dependency on neighboring PEs, which enforces the need for data communication among PEs. These point-to-point communications are performed from the PE boundary points into the corresponding ghost/buffer points of neighboring PEs and are locally synchronized at every stage of the time integration.

At extreme scales, the communication and synchronization of data incur significant overheads, posing a major bottleneck to scalability, and rendering the standard synchronous approach prohibitively expensive. To minimize these communication overheads, several efforts have been made at different levels across hardware and software stacks \cite{ecp-codesign}. At a mathematical level, different asynchronous approaches for partial differential equation (PDE) solvers were introduced in \cite{Amitai_CMA_1992,Amitai_NA_1994,Aditya_SC_2012,Donzis_JCP_2014,Mittal_PRE_2017}, where data synchronization between PEs is relaxed, and the solvers are allowed to proceed with computations using delayed data at PE boundaries regardless of the status of communications. The asynchronous finite difference method proposed in \cite{Aditya_SC_2012,Donzis_JCP_2014} provides a general framework to analyze numerical properties with asynchrony, and shows that standard finite difference schemes are at most first-order accurate when delayed values are used in stencils. This issue was addressed by the development of asynchrony-tolerant (AT) schemes that use wider spatio-temporal stencils at PE boundaries and provide arbitrary orders of accuracy \cite{Aditya_JCP_2017}. The asynchronous finite difference method can be implemented in solvers based on two parallel algorithms: first, the communication avoiding algorithm (CAA), which relaxes communications over regular intervals during time advancement, and second, the synchronization avoiding algorithm (SAA), which initiates communications at every time step but does not enforce explicit synchronization of data \cite{Aditya_arxiv_2019,Komal_JCP_2020}. The efficacy and scalability of these algorithms have been demonstrated using canonical problems, such as Burgers' turbulence \cite{Aditya_arxiv_2019,Shubham_JCP_2023} and compressible isotropic turbulence \cite{Komal_JCP_2020}. For instance, asynchronous algorithms achieved a maximum speed-up of $6.68\times$ relative to the synchronous algorithm at an extreme scale of 27,000 cores \cite{Shubham_JCP_2023}. It should also be noted that when multistage time integration schemes, such as low-storage explicit Runge-Kutta (LSERK), are used, asynchronous algorithms can eliminate communication completely in the intermediate stages \cite{Shubham_JCP_2023}. Furthermore, the numerical accuracy of the asynchronous finite difference method has been assessed on one-dimensional chemically reacting flows \cite{Komal_JCP_2023}. The asynchronous approach has also been extended to the discontinuous Galerkin (DG) method \cite{Shubham_AIAA_2022} and used to solve compressible Euler equations \cite{Shubham_CMAME_2024} and reacting flow equations \cite{Aswin_arxiv_2025}.

The errors incurred at the PE boundaries due to the use of delayed data were observed to introduce high-frequency numerical oscillations. These numerical oscillations can potentially alter the flow structures and important features of transitional and turbulent flows. In this regard, the efficacy of high-order AT schemes in restricting these numerical errors to insignificant amounts must be investigated. To this end, the two asynchronous algorithms are implemented in the high-order compressible flow solver COMP-SQUARE to investigate the accuracy and scalability of the asynchronous approach on sensitive test cases involving transition to turbulence. COMP-SQUARE is a well established code that solves a wide range of practically relevant flow problems in complex geometries, such as transitional and turbulent boundary layers \cite{Rao_FTC_2018,Ananth_JoT_2023}, separated flows in aircraft engine intakes \cite{Rao_DLES_2019,Tyacke_PAS_2019,Adrian_AIAA_2024}, airfoils \cite{Lin_TAJ_2017} and cavities \cite{Ganesh_FTC_2022}. The remainder of this paper is organized as follows. Solver details, including the governing equations, standard numerical schemes, and parallelization method, are provided in Sec.~\ref{sec:math}. The concept of asynchronous computing is described in Sec.~\ref{sec:async_math} followed by the parallel implementation of asynchronous solvers in Sec.~\ref{sec:parallel_implementation}. The numerical experiments and results from three test cases, isentropic vortex advection, Taylor-Green vortex, and transitional flow over an airfoil, are detailed in Sec.~\ref{sec:results}. Finally, the conclusions are presented in Sec.~\ref{sec:conclusions}.

\section{COMP-SQUARE solver details\label{sec:math}}
The high-order structured multiblock compressible flow solver COMP-SQUARE solves the unsteady non-dimensional mass, momentum, and energy conservation equations on the general curvilinear coordinates given by Eq.~\ref{eq:NS_eqn}. Here $\bm{U} = \{\rho,\rho u, \rho v, \rho w, \rho E_t\}$ is the vector of the conserved variables and $J = \partial (\xi, \eta, \zeta)/\partial (x, y, z)$ is the coordinate transformation Jacobian, $\rho$ is the density, $u$, $v$ and $w$ are the components of the velocity, and $E_t$ is the total specific energy. The advective fluxes $\bm{\hat{F}}$, $\bm{\hat{G}}$, $\bm{\hat{H}}$ and the viscous fluxes $\bm{\hat{F}}_{v}$, $\bm{\hat{G}}_{v}$, $\bm{\hat{H}}_{v}$ are defined in the Appendix~\ref{sec:flux_math} for completeness.
\begin{equation}     
\frac{\partial}{\partial t}\left(\frac{\bm{U}}{J}\right) + \frac{\partial \bm{\hat{F}}}{\partial \xi} + \frac{\partial \bm{\hat{G}}}{\partial \eta} + \frac{\partial \bm{\hat{H}}}{\partial \zeta} = \frac{1}{Re}\left[\frac{\partial \bm{\hat{F}}_{v}}{\partial \xi} + \frac{\partial \bm{\hat{G}}_{v}}{\partial \eta} + \frac{\partial \bm{\hat{H}}_{v}}{\partial \zeta}\right]     
\label{eq:NS_eqn} 
\end{equation}
The solver computes spatial derivatives using high-order explicit and compact (up to sixth-order accurate) finite-difference schemes. Specifically, the spatial derivative $\phi^{'}$ of a general function $\phi$ is computed using the generalized stencil 
\begin{equation}     
\alpha \phi^{'}_{i-1} + \phi^{'}_{i} + \alpha \phi^{'}_{i+1} = b\frac{\phi_{i+2}-\phi_{i-2}}{4\Delta\xi} + a\frac{\phi_{i+1}-\phi_{i-1}}{2\Delta\xi},     
\label{eq:spat_disc} 
\end{equation} 
where $\alpha = 0$ for explicit schemes \cite{Lele_JCP_1992}, and $a$ and $b$ determine the order of accuracy. High-frequency numerical oscillations that arise due to nonlinear interactions in the solution are suppressed using high-order Pad\'e type low-pass filters (up to tenth-order accuracy) with a filter coefficient $\alpha_{f}$ where $-0.5 < \alpha_{f} \leq 0.5$ \cite{Visbal_JCP_2002}. The solver is equipped with classical fourth-order Runge-Kutta and low-storage explicit Runge-Kutta (LSERK) schemes for time integration \cite{Williamson_JCP_1980,Kennedy_ANM_2000,Carpenter_report_1994}. For an $s$ stage LSERK method, the equations to compute $\bm{U}^{n+1}$ from $\bm{U}^{n}$ are given in Eq.~\ref{eq:lserk_Q}, where $\bm{Q}^{(m)}$ is the vector of intermediate variables, $A_m$ and $B_m$ are the coefficients at the $m$th stage, where $m=1,2,...,s$ such that $\bm{U}^{(0)}\equiv\bm{U}^{n}$ and $\bm{U}^{(s)}\equiv\bm{U}^{n+1}$. The values of the coefficients used in this study are presented in Appendix~\ref{app:fd-schemes}.
\begin{equation} 
\begin{split}     
\bm{Q}^{(m)} &= A_m\bm{Q}^{(m-1)} + \Delta t \bm{R}(\bm{U}^{(m-1)})\\     \bm{U}^{(m)} &= \bm{U}^{(m-1)} + B_m \bm{Q}^{(m)} 
\end{split} 
\label{eq:lserk_Q} 
\end{equation} 

The solver performs multiblock decomposition of the computational domain to preserve high fidelity and is especially important in problems involving complex geometries, such as turbine blades \cite{Ananth_JoT_2023} and engine intakes \cite{Adrian_AIAA_2024}. The code is parallelized to scale on multiple central processing units (CPUs) using non-blocking communication subroutines from the message passing interface (MPI) framework. OpenACC directives are also incorporated to accelerate computations on graphical processing units (GPUs). Nevertheless, all simulations reported in this paper exploit the multi-CPU version of the solver.

\section{Asynchronous computing approach\label{sec:async_math}}
In parallel DNS implementations, the computational domain is decomposed into several subdomains that are mapped to several processing elements (PEs). The physical grid points in each subdomain can be classified into two groups based on the data dependency on communication: interior points and PE boundary points. The spatial derivatives at the interior points are computed using finite-difference stencils that contain physical grid points within the PE. On the other hand, the stencils for computing spatial derivatives at PE boundary points include grid points from other PEs, requiring data movement among PEs. Data communication is performed by sending solution values from PE boundary points and copying them into the ghost/buffer points of neighboring PEs. These communications are initiated between PEs and explicitly synchronized at every stage of time integration, referring to this procedure as the standard \textit{synchronous approach}. At extreme scales, such communications and synchronizations result in significant overheads and, in turn, severely degrade the parallel performance.

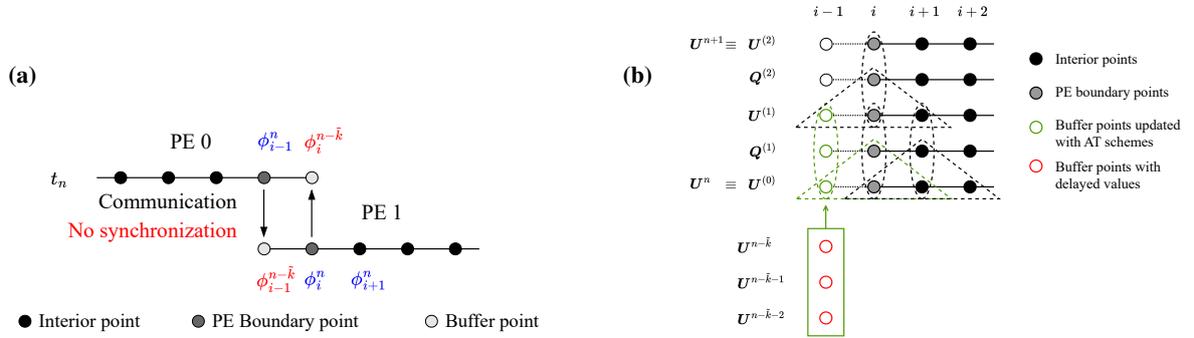
\begin{figure}[h!]
\centering
\includegraphics[width=8cm]{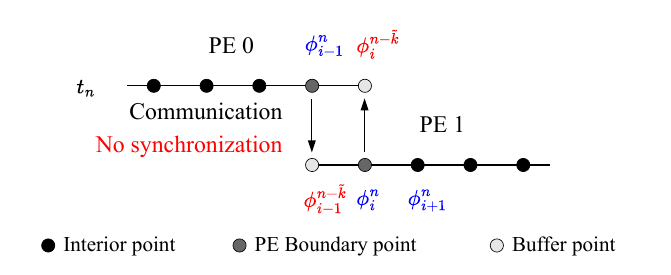}
\includegraphics[width=8cm]{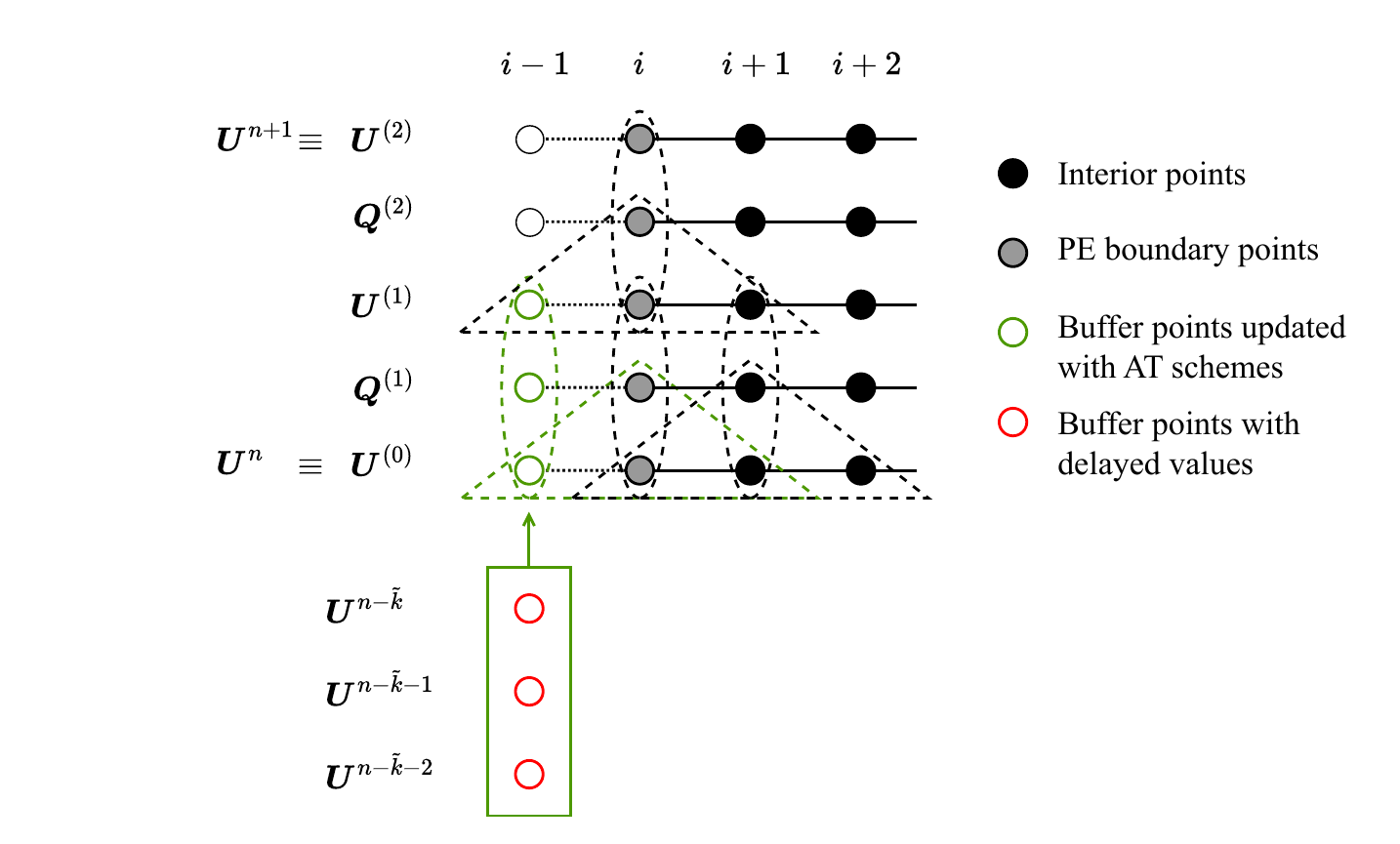}
\begin{picture}(0,0)
\put(-450,100){\bf \small (a)}
\put(-220,100){\bf \small (b)}
\end{picture}
\caption{(a) An illustration of communications with relaxed synchronization at the interface between two processing elements (PEs). The delay in time levels at the buffer points is $\tilde{k}$. (b) Pattern of LSERK2-CD2-AT2 stencil computations at a left PE boundary point.}
\label{fig:async_schematic}
\end{figure}
Based on the asynchronous finite-difference method, the overhead associated with data communications can be reduced either by relaxing their synchronization or by periodically avoiding communication for some time steps. Thus, the solver can potentially perform computations using delayed/stale data in buffer points at the PE boundaries. In Fig.~\ref{fig:async_schematic}(a), $\phi_{i}^{n}$ is the function value at the left boundary point $i$ of PE 1 at time level $n$, whereas $\phi_{i}^{n-\tilde{k}}$ is the function value at the buffer point of PE 0 at time level $n-\tilde{k}$. $\tilde{k}$ is the delay, where $(\tilde{.})$ denotes its random nature due to factors like performance variation and network topology in the parallel setting. Such an asynchronous approach was shown to be numerically consistent and stable, but only exhibits first-order accuracy irrespective of the spatial order of the standard finite difference schemes \cite{Aditya_SC_2012,Donzis_JCP_2014}. In a subsequent study, asynchrony-tolerant (AT) schemes for computations at PE boundaries were proposed, which preserve high-order accuracy while operating on delayed data in buffers \cite{Aditya_JCP_2017}. Using these schemes, the spatial derivatives at the PE boundary points are approximated based on extended spatio-temporal stencils consisting of multiple delayed time levels at buffer points. Furthermore, when multistage Runge-Kutta (RK) schemes are used for time integration, communications/synchronizations are relaxed over the intermediate stages \cite{Shubham_JCP_2023}. This requires additional asynchrony-tolerant corrections at the buffer points to account for the fractional time-step advancement between RK stages. It should be noted that these additional computations using AT schemes are performed only at PE boundaries, which are typically a small fraction of the total grid points in a subdomain. Standard finite difference schemes are used at interior points (Eq.~\ref{eq:spat_disc}). Before we outline the procedure for implementing the asynchronous method, the necessary AT schemes that are used near the PE boundaries are described in the following sections.

\subsection{Asynchrony-tolerant (AT) schemes\label{subsec:at_schemes}}
\noindent \textbf{PE boundary points:}
Let us consider the $d$th spatial derivative of a function $\phi(\xi,t)$ expressed in a general spatio-temporal stencil
\begin{equation}
    \frac{\partial^d \phi}{\partial \xi^d}\Biggr|_{i}^{n} = \sum_{j=-J_1}^{J_2} \sum_{l=0}^{L-1} \tilde{c}_{j}^{l} \phi_{i+j}^{n-l} + \mathcal{O}(\Delta \xi^{q}),
    \label{eq:at_spat_gen}
\end{equation}
where $j \in \{-J_1,..,J_2\}$, $J_1$ and $J_2$ are the extents of the spatial stencil to the left and right of point $i$, respectively, and $l \in \{0,..,L-1\}$ is the extent of the temporal stencil with delayed time levels $n-l$ bounded by a maximum allowable delay $L$. $\mathcal{O}(\Delta \xi^{q})$ denotes that the approximation is $q$th-order accurate in space. The Taylor series expansion for $\phi_{i+j}^{n-l}$ with respect to $\phi_{i}^{n}$ on a general spatio-temporal stencil can be written as
\begin{equation}
    \phi_{i+j}^{n-l} = \sum_{\lambda=0}^{\infty}\sum_{\theta=0}^{\infty} \phi^{(\lambda,\theta)} \frac{(j\Delta \xi)^{\lambda}(-l\Delta t)^{\theta}}{\lambda ! \theta !},
    \label{eq:spat_temp_taylor}
\end{equation}
where $\phi^{(\lambda,\theta)}$ denotes the partial derivative of $\phi_{i}^{n}$ of order $\lambda$ and $\theta$ along $\xi$ and $t$, respectively. In order to compute the spatial derivative based on such a hybrid stencil, necessary lower order terms must be eliminated upon substituting the Taylor series in Eq.~\ref{eq:spat_temp_taylor} into the linear combination in Eq.~\ref{eq:at_spat_gen}. The lower order terms depend on the desired order of accuracy and the scaling exponent $r$ in the stability criterion $(\Delta t \sim \Delta \xi^{r})$. The time integration scheme must be at least $\mathcal{O}(\Delta t^{q/r})$ accurate to obtain an overall accuracy of $\mathcal{O}(\Delta \xi^{q})$ in space. The coefficients $\tilde{c}_{j}^{l}$ in Eq.~\ref{eq:at_spat_gen} can be determined by retaining the $d$th-order derivative term and eliminating appropriate lower order terms with the constraints
\begin{equation}
    \sum_{j=-J_1}^{J_2} \sum_{l=0}^{L-1} \tilde{c}_{j}^{l} \frac{(j\Delta \xi)^{\lambda}(-l\Delta t)^{\theta}}{\lambda ! \theta !} = 
    \begin{cases}
    1 & \text{for\ } (\lambda,\theta) = (d,0) \\
    0 & \text{for\ } \lambda + r\theta < d + q; (\lambda,\theta) \neq (d,0).
    \end{cases}
    \label{eq:at_spat_constraints}
\end{equation}
These constraints result in a linear system of equations to solve for the unknown coefficients $\tilde{c}^{l}_{j}$ that correspond to delayed time levels \{$n-\tilde{k},n-\tilde{k}-1,\cdots,n-\tilde{k}-C+1$\}, where $\tilde{k}$ and $C$ denote the latest delay and the number of time levels, respectively. For example, consider the schematic of the stencil at the left boundary of a PE shown in Fig.~\ref{fig:async_schematic}(b). Let us approximate the first-order derivative using a second-order accurate scheme at the left PE boundary point $i$. Stability criterion based on the advective Courant-Friedrichs-Lewy (CFL) $\Delta t \sim \Delta \xi\ (r=1)$ is used. The unknown coefficients $\tilde{c}^{l}_{j}$ can be determined by solving Eq.~\ref{eq:at_spat_constraints} based on the required parameters $(d,q,r)$. The spatial derivative at the zeroth stage at time level $n$ at the left PE boundary point $i$ (denoted by gray filled circle) is computed based on a stencil that contains the solutions from time levels $n-\tilde{k}$, $n-\tilde{k}-1$ and $n-\tilde{k}-2$ at the buffer point $i-1$ (denoted by white circle). The AT scheme for the spatial derivative of $\phi$ at the grid point $i$ can be approximated in the general form
\begin{equation}
    \frac{\partial \phi}{\partial \xi}\Biggr|_{i}^{n} \approx \boldsymbol{\tilde{c}}(\tilde{k})^{T}\boldsymbol{\Phi}(\tilde{k}),
    \label{eq:at_spat_matmul}
\end{equation}
where $\boldsymbol{\tilde{c}}^{}(\tilde{k})=[\tilde{c}^{\tilde{k}}, \tilde{c}^{\tilde{k}+1},\ldots, \tilde{c}^{\tilde{k}+C-1}]^{T}$ is the vector containing the unknown coefficients and $\boldsymbol{\Phi}(\tilde{k})$ is the vector of spatial derivatives of $\phi$ with delayed data at buffer points. For the parameters $(d,q,r)=(1,2,1)$, the second-order asynchrony-tolerant (AT2) scheme corresponds to
\begin{equation*}
\boldsymbol{\tilde{c}}(\tilde{k}) =    \begin{bmatrix}
\dfrac{1}{2}\left(\tilde{k}^2 + 3\tilde{k} + 2\right)\\
-\left(\tilde{k}^2 + 2\tilde{k}\right)\\
\dfrac{1}{2}\left(\tilde{k}^2 + \tilde{k}\right)
\end{bmatrix},\
\boldsymbol{\Phi}(\tilde{k}) =    \begin{bmatrix}
a\dfrac{\left(\phi_{i+1}^{n}-\phi_{i-1}^{n-\tilde{k}}\right)}{2\Delta\xi}\\
a\dfrac{\left(\phi_{i+1}^{n}-\phi_{i-1}^{n-\tilde{k}-1}\right)}{2\Delta\xi}\\
a\dfrac{\left(\phi_{i+1}^{n}-\phi_{i-1}^{n-\tilde{k}-2}\right)}{2\Delta\xi}
\end{bmatrix}.
\label{eq:at_left_o2}
\end{equation*}
As illustrated in Fig.~\ref{fig:async_schematic}(b), the spatial derivatives of flux functions at the left PE boundary point $i$ are computed using solutions from multiple delayed time levels at buffer point $i-1$. Similarly, for a fourth-order central difference approximation, the parameters are $(d,q,r)=(1,4,1)$. The AT4 scheme for the spatial derivative at the left PE boundary point can be expressed in the same form as Eq.~\ref{eq:at_spat_matmul}, with
\begin{equation*}
\boldsymbol{\tilde{c}}(\tilde{k}) =    \begin{bmatrix}
\dfrac{1}{24}\left(\tilde{k}^4 + 10\tilde{k}^3 + 35\tilde{k}^2 + 50\tilde{k} + 24\right)\\
- \dfrac{1}{6}\left(\tilde{k}^4 + 9\tilde{k}^3 + 25\tilde{k}^2 + 24\tilde{k}\right)\\
\dfrac{1}{4}\left(\tilde{k}^4 + 8\tilde{k}^3 + 19\tilde{k}^2 + 12\tilde{k}\right)\\
- \dfrac{1}{6}\left(\tilde{k}^4 + 7\tilde{k}^3 + 14\tilde{k}^2 + 8\tilde{k}\right)\\
\dfrac{1}{24}\left(\tilde{k}^4 + 6\tilde{k}^3 + 11\tilde{k}^2 + 6\tilde{k}\right)
\end{bmatrix},\
\boldsymbol{\Phi}(\tilde{k}) =    \begin{bmatrix}
b\dfrac{\phi_{i+2}^{n}-\phi_{i-2}^{n-\tilde{k}}}{4\Delta\xi} + a\dfrac{\phi_{i+1}^{n}-\phi_{i-1}^{n-\tilde{k}}}{2\Delta\xi}\\
b\dfrac{\phi_{i+2}^{n}-\phi_{i-2}^{n-\tilde{k}-1}}{4\Delta\xi} + a\dfrac{\phi_{i+1}^{n}-\phi_{i-1}^{n-\tilde{k}-1}}{2\Delta\xi}\\
b\dfrac{\phi_{i+2}^{n}-\phi_{i-2}^{n-\tilde{k}-2}}{4\Delta\xi} + a\dfrac{\phi_{i+1}^{n}-\phi_{i-1}^{n-\tilde{k}-2}}{2\Delta\xi}\\
b\dfrac{\phi_{i+2}^{n}-\phi_{i-2}^{n-\tilde{k}-3}}{4\Delta\xi} + a\dfrac{\phi_{i+1}^{n}-\phi_{i-1}^{n-\tilde{k}-3}}{2\Delta\xi}\\
b\dfrac{\phi_{i+2}^{n}-\phi_{i-2}^{n-\tilde{k}-4}}{4\Delta\xi} + a\dfrac{\phi_{i+1}^{n}-\phi_{i-1}^{n-\tilde{k}-4}}{2\Delta\xi}
\end{bmatrix}.
\label{eq:at_left_o4}
\end{equation*}
We note that the coefficient vector $\boldsymbol{\tilde{c}}(\tilde{k})$ remains the same for the right PE boundary point, while $\boldsymbol{\Phi}(\tilde{k})$ is modified such that $i+1$ and $i+2$ are buffer points with delays whereas $i-1$ and $i-2$ are interior points. We also note that the spatial derivatives at PE boundary points obtained using AT schemes are based on an extended stencil in time alone, while the spatial extent remains the same relative to the standard central difference schemes. In general, the stencil could be expanded in both space and time. It should also be noted that introducing delays at PE boundaries gives rise to high frequency numerical oscillations, leading to stricter CFL conditions. However, the use of AT schemes improves the stability criterion to some extent compared to using standard schemes in the presence of delays \cite{Komal_JCP_2020}.

\vspace{10pt}
\noindent\textbf{Buffer points:} In the absence of communications among PEs in the intermediate LSERK stages, asynchrony-tolerant (AT) schemes can be implemented at the buffer points in a naive manner by considering the fractional time step advancement at each RK stage in addition to the delayed levels across time steps (see Fig.~\ref{fig:async_schematic}(b) for reference). However, this naive approach leads to poor accuracy (at most third-order accurate) as previously studied \cite{Shubham_JCP_2023}. To overcome this issue, buffer points are updated at each RK stage using the delayed solution values by replacing the spatial derivative in Eq.~\ref{eq:lserk_Q} with a temporal derivative as shown in Eq.~\ref{eq:lserk_at_buffer}.
\begin{equation}
\begin{aligned}
    \bm{Q}_{\text{buf}}^{(m)} &= A_m\bm{Q}_{\text{buf}}^{(m-1)} + \Delta t \bm{R}_{\text{buf}}^{(m-1)},~~~~ m = 1,2,...,s\\
    &= A_m\bm{Q}_{\text{buf}}^{(m-1)} + \Delta t \frac{\partial \bm{U}}{\partial t}\Biggr|_{\text{buf}}^{(m-1)}.
\end{aligned}
\label{eq:lserk_at_buffer}
\end{equation}
The temporal derivative in Eq.~\ref{eq:lserk_at_buffer} is to be computed at every RK stage using delayed solution values such that the necessary lower order terms are eliminated to retain higher order accuracy. The temporal derivative of a function $\phi$ at stage $e\equiv m-1$ and at grid point $i$ can be written in terms of function values at previous time levels as expressed in Eq.~\ref{eq:buffer_temp_deriv}.
\begin{equation}
    \frac{\partial \phi}{\partial t}\Biggr|_{i}^{(e)} = \sum_{l=L_1}^{L_2} \tilde{c}^{l} \phi_{i}^{n-l},~
    \phi_{i}^{n-l} = \sum_{\theta=0}^{\infty} \frac{(-l\Delta t)^{\theta}}{\theta !} \frac{\partial^{\theta}\phi}{\partial t^{\theta}}\Biggr|_{i}^{n}.
    \label{eq:buffer_temp_deriv}
\end{equation}
The Taylor series expansion of the LSERK scheme is expressed in Eq.~\ref{eq:lserk_expans_gen}, where $\nu_{0}^{e} = 1$, $\nu_{1}^{e} = \sum\limits_{j=1}^{e}B_j + \sum\limits_{j=2}^{e}A_jB_j + \sum\limits_{j=2}^{e-1}A_jA_{j+1}B_{j+1} + ... + B_e\prod\limits_{j=2}^{e}A_j$, ..., and $\nu_{e}^{e} = \prod\limits_{j=1}^{e}B_j$.
\begin{equation}
    \phi_{i}^{(e)} = \nu_{0}^{e}\phi_{i}^{n} + \nu_{1}^{e}\Delta t \frac{\partial \phi}{\partial t}\Biggr|_{i}^{n} + ... + \nu_{e}^{e} (\Delta t)^{e}\frac{\partial^e \phi}{\partial t^e}\Biggr|_{i}^{n}.
    \label{eq:lserk_expans_gen}
\end{equation}
The derivative of Eq.~\ref{eq:lserk_expans_gen} with respect to time is
\begin{equation}
    \frac{\partial \phi}{\partial t}\Biggr|_{i}^{(e)} = \nu_{0}^{e}\frac{\partial\phi}{\partial t}\Biggr|_{i}^{n} + \nu_{1}^{e}\Delta t \frac{\partial^2 \phi}{\partial t^2}\Biggr|_{i}^{n} + ... + \nu_{e}^{e} (\Delta t)^{e}\frac{\partial^{e+1} \phi}{\partial t^{e+1}}\Biggr|_{i}^{n}.
    \label{eq:time_deriv_expans}
\end{equation}
Equations~\ref{eq:buffer_temp_deriv} and~\ref{eq:time_deriv_expans} are equivalent and the coefficients are compared, giving the necessary constraints to satisfy $\mathcal{O}(\Delta \xi^{q})$ accuracy, as shown in Eq.~\ref{eq:at_buffer_constraints}.
\begin{equation}
    \sum_{j=0}^{e}\nu_{j}^{e}(\Delta t)^{j}\frac{\partial^{j+1} \phi}{\partial t^{j+1}} = \sum_{l=L_1}^{L_2}\tilde{c}^{l}\sum_{\theta=0}^{\infty}\frac{(-l\Delta t)^{\theta}}{\theta !} \frac{\partial^{\theta}\phi}{\partial t^{\theta}},
    \label{eq:at_buffer_taylor}
\end{equation}
\begin{equation}
    \sum_{l=L_1}^{L_2} \tilde{c}^{l} \frac{(-l\Delta t)^{\theta}}{\theta !} = 
    \begin{cases}
       \nu_{j}^{e}(\Delta t)^{j} & \text{for\ }\theta=j+1,\ j=0,...,e\\
        0 & \text{for\ }\theta = 0\ \text{and\ } e + 1 < \theta < \dfrac{q}{r} + 1.
    \end{cases}
    \label{eq:at_buffer_constraints}
\end{equation}
\\

\noindent\textbf{Buffer update schemes:} Equation~\ref{eq:at_buffer_constraints} is solved for a given pair $(q,r)$ to obtain the coefficients of the AT buffer update schemes at the intermediate stages of the LSERK method. The formulae for computing the temporal derivative of $\phi$ are listed in Appendix~\ref{sec:at_buffer_appendix} for the two-stage LSERK2 and the five-stage LSERK4 methods. We note that buffer updates are performed until the $(s-1)$th stage which corresponds to $e=s-2$.\\

\noindent\textbf{Buffer extrapolation schemes:}
At the zeroth stage of the LSERK method, the solution at a buffer point is approximated as a polynomial extrapolation of solutions from multiple delayed time levels. The second-order accurate extrapolation scheme for $\phi_{i}^{(0)}$ in terms of the delayed function values is
\begin{equation}
    \phi_{i}^{(0)} = \frac{1}{2}\left(\tilde{k}^2 + 3\tilde{k} + 2\right)\phi_{i}^{n-\tilde{k}} - \left(\tilde{k}^2 + 2\tilde{k}\right)\phi_{i}^{n-\tilde{k}-1} + \frac{1}{2}\left(\tilde{k}^2 + \tilde{k}\right)\phi_{i}^{n-\tilde{k}-2}. \\
\label{eq:buffer_extrap_o2}
\end{equation}
Similarly, the fourth-order accurate extrapolation scheme for $\phi_{i}^{(0)}$ is
\begin{equation}
\begin{aligned}
    \phi_{i}^{(0)} =\ & \frac{1}{24}\left(\tilde{k}^4 + 10\tilde{k}^3 + 35\tilde{k}^2 + 50\tilde{k} + 24\right)\phi_{i}^{n-\tilde{k}} - \frac{1}{6}\left(\tilde{k}^4 + 9\tilde{k}^3 + 25\tilde{k}^2 + 24\tilde{k}\right)\phi_{i}^{n-\tilde{k}-1} \\
    & + \frac{1}{4}\left(\tilde{k}^4 + 8\tilde{k}^3 + 19\tilde{k}^2 + 12\tilde{k}\right)\phi_{i}^{n-\tilde{k}-2} - \frac{1}{6}\left(\tilde{k}^4 + 7\tilde{k}^3 + 14\tilde{k}^2 + 8\tilde{k}\right)\phi_{i}^{n-\tilde{k}-3} \\
    & + \frac{1}{24}\left(\tilde{k}^4 + 6\tilde{k}^3 + 11\tilde{k}^2 + 6\tilde{k}\right)\phi_{i}^{n-\tilde{k}-4}. \\
\end{aligned}
\label{eq:buffer_extrap_o4}
\end{equation}

\subsection{Implementation of AT schemes\label{subsec:at_implementation}}
In this section, we explain the implementation of the LSERK-AT method that provides $\mathcal{O}(\Delta \xi^{q})$ accurate solutions in space. Let us consider the schematic of the LSERK2-CD2-AT2 method shown in Fig.~\ref{fig:async_schematic}(b) for reference. CD2 denotes the standard second-order central difference scheme used at interior points while AT2 denotes the second-order asynchrony-tolerant schemes used at PE boundaries. Here, the interior points (denoted by black, filled circles) are independent of communications between PEs and therefore the spatial derivatives at these points can be computed using standard finite difference methods. The computation of spatial derivatives at PE boundary points (denoted by gray, filled circles) does indeed depend on data from neighboring PEs which is stored at the buffer points (denoted by white circles). In the first stage of the LSERK-AT method, the PE boundary points are updated using AT schemes. As discussed in Sec.~\ref{subsec:at_schemes}, Eq.~\ref{eq:at_spat_matmul} can be used to approximate the spatial derivative based on delayed data at buffer points. Meanwhile, the solution values $\bm{U}^{(0)}$ at buffer points are approximated to the current fractional time step with a polynomial extrapolation using delayed buffer data. In the subsequent stages, the buffer points are updated based on the modified update scheme in Eq.~\ref{eq:lserk_at_buffer} that replaces spatial derivatives with temporal derivatives. These update schemes are specific to the LSERK scheme used, as they selectively eliminate lower order terms from the Taylor series such that the solution is $\mathcal{O}(\Delta t^{q/r})$ accurate after $s$ stages. Since the buffer points are updated using this accurate approach in the intermediate stages, the spatial derivatives at the PE boundary points are computed using standard finite difference schemes. We note that for an $s$ stage LSERK scheme, the AT buffer updates are performed until the $(s-1)$th stage. The asynchronous solver requires additional memory to store the multi-level buffers of size $C\times N_{\text{buf}}$ where $C$ and $N_{\text{buf}}$ denote the number of delayed time levels and number of buffer points in the subdomain, respectively. It may appear that performing these AT corrections introduces additional computational overhead, however, these computations are fairly localized and exhibit high arithmetic intensity. In previous studies, the asynchronous solvers are observed to incur a small fraction of computational overhead but provide significant gains in communications \cite{Shubham_JCP_2023}.

\section{Parallel implementation\label{sec:parallel_implementation}}

The baseline synchronous algorithm is first described in this section for reference. The relaxation in communication/synchronization using asynchronous computing approach can be realized using two parallel algorithms. First, the communication avoiding algorithm, where communications are performed in a synchronous manner but once every few time steps. Second, the synchronization avoiding algorithm, where data communications are initiated but the explicit synchronization is not enforced.

\newcommand\SAalg{
\For{$m=1,...,s$}
\State $\bm{U}^{(m)} = f(\bm{U}^{(m-1)})$
\State $\bm{U}_{\text{send}} \leftarrow \bm{U}_{\text{bnd}}$
\State Initiate \texttt{MPI\_Isend} and \texttt{MPI\_Irecv} requests
\State Call blocking \texttt{MPI\_Wait}
\State $\bm{U}_{\text{buf}} \leftarrow \bm{U}_{\text{recv}}$
\EndFor
}

\subsection{Synchronous algorithm (SA)\label{subsec:sa}}

In the standard synchronous algorithm (SA) based on Runge-Kutta time integration schemes, an additional stage loop within the main time loop is incorporated to advance the RK stages. As described in Algorithm~\ref{alg:sa}, upon updating the solution at each RK stage,  data communication between PEs is performed, where the solutions at the PE boundary points are copied to a send buffer and a send request is initiated using the \texttt{MPI\_Isend} subroutine. Meanwhile, the neighboring PE initiates the receive request using the \texttt{MPI\_Irecv} subroutine. The synchronization of communications is ensured by calling the blocking \texttt{MPI\_Wait} subroutine, and the received data are copied to the buffer points. The algorithm proceeds to the next RK stage of time advancement only after all buffer points are updated. An additional optimization for communication overheads in this synchronous algorithm can include the overlap between the computations at the interior points and communications near the PE boundaries.

\begin{figure}[H]
\begin{algorithm}[H]
\caption{Synchronous algorithm (SA)}
\begin{algorithmic}[1]

\For{$n=1,...,nsteps$}
\SAalg
\EndFor

\end{algorithmic}
\label{alg:sa}
\end{algorithm}
\end{figure}

\subsection{Communication avoiding algorithm (CAA)\label{subsec:caa}}

The communication avoiding algorithm (CAA) reduces data movement among PEs by skipping communication for a finite number of time steps over regular intervals. As explained in Algorithm~\ref{alg:caa}, the synchronous algorithm is carried out for the first $C$ time steps to populate the buffer points with data of $C$ time levels, followed by the asynchronous algorithm for the remaining time steps. The spatial derivatives at the PE boundaries are computed using the AT schemes and multi-level buffer data. In the CAA, communications are performed based on the condition associated with $n\%L$ where $n$ and $L$ denote the $n$th time step and the maximum allowable delay, respectively. At every time step, the latest delay is incremented by one level ($\tilde{k}\leftarrow \tilde{k}+1$) across all PEs. If $0 \leq n\%L \leq C-1$, the solutions at the PE boundary points are copied to an auxiliary send buffer with dimension $C$ in time. If $n\%L == C-1$, the auxiliary buffer data are communicated to the receiving buffers, and the delay is reset ($\tilde{k}\leftarrow 0$). This results in uniform and periodic behavior of delays across all PEs.

\begin{figure}[h]
\begin{algorithm}[H]
\caption{Communication avoiding algorithm (CAA)}
\begin{algorithmic}[1]
\For{$n=1,...,C$} \Comment{\textbf{Synchronous loop for first $C$ steps}}
\SAalg
\EndFor

\For{$n=C+1,...,nsteps$} \Comment{\textbf{Asynchronous loop}}
\For{$m=1,...,s$}
\State $\bm{U}^{(m)}_{\text{buf}} = f(\bm{U}^{(m-1)}_{\text{buf}},\bm{U}^{n-\tilde{k}}_{\text{buf}},\cdots,\bm{U}^{n-\tilde{k}-C+1}_{\text{buf}})$
\State $\bm{U}^{(m)} = f(\bm{U}^{(m-1)})$
\EndFor
\State Update delay: $\tilde{k} = \tilde{k} + 1$
\If{$0 \leq n\%L \leq C-1$}
\State $\bm{U}_{\text{send}} \leftarrow \bm{U}_{\text{bnd}}$
\EndIf
\If{$n\%L == C-1$}
\State Initiate \texttt{MPI\_Isend} and \texttt{MPI\_Irecv} requests
\State Call blocking \texttt{MPI\_Wait}
\State $\bm{U}_{\text{buf}} \leftarrow \bm{U}_{\text{recv}}$
\State Reset delay: $\tilde{k} = 0$
\EndIf
\EndFor

\end{algorithmic}
\label{alg:caa}
\end{algorithm}
\end{figure}

\subsection{Synchronization avoiding algorithm (SAA)\label{subsec:saa}}

The synchronization avoiding algorithm (SAA) relaxes the communication overhead by not enforcing synchronization, such that the solver proceeds with computations regardless of the status of their completion. As shown in Algorithm~\ref{alg:saa}, the first $C$ time steps are executed in a synchronous manner to populate the buffer points, followed by the asynchronous loop, where non-blocking \texttt{MPI\_Isend} and \texttt{MPI\_Irecv} requests are initiated at every time step. The status of the communication requests and the latest delay are obtained using the \texttt{MPI\_Test} subroutine. If the latest delay reaches the maximum allowable value $L$ or if the buffer point data are not from consecutive time levels, all existing communications are locally synchronized using \texttt{MPI\_Wait}, and the delay is reset. The behavior of delays in SAA depends on several factors, such as workload balancing among PEs, the network topology of the architecture, and the communication library used. These factors result in stochastic and non-uniform behavior of delays across PEs.

\begin{figure}[H]
\begin{algorithm}[H]
\caption{Synchronization avoiding algorithm (SAA)}
\begin{algorithmic}[1]

\For{$n=1,...,C$} \Comment{\textbf{Synchronous loop for first $C$ steps}}
\SAalg
\EndFor

\For{$n=C+1,...,nsteps$} \Comment{\textbf{Asynchronous loop}}
\For{$m=1,...,s$}
\State $\bm{U}^{(m)}_{\text{buf}} = f(\bm{U}^{(m-1)}_{\text{buf}},\bm{U}^{n-\tilde{k}}_{\text{buf}},\cdots,\bm{U}^{n-\tilde{k}-C+1}_{\text{buf}})$
\State $\bm{U}^{(m)} = f(\bm{U}^{(m-1)})$
\EndFor
\State Initiate \texttt{MPI\_Isend} and \texttt{MPI\_Irecv} requests
\State Check communication status using \texttt{MPI\_Test} and get latest delay $\tilde{k}$
\If{$\tilde{k} > L-1$ OR time levels not consecutive}
\State Synchronize communication using \texttt{MPI\_Wait}
\State Reset delay $\tilde{k} = 0$
\EndIf
\EndFor

\end{algorithmic}
\label{alg:saa}
\end{algorithm}
\end{figure}

The performance evaluations of the above parallel algorithms are carried out on two supercomputers. First, PARAM Pravega at SERC, IISc where each compute node is built using two Intel Xeon Cascade 8268 2.9 GHz processors with 24 cores each and contains 192 GB RAM, connected via Infiniband interconnect. Second, PARAM Rudra at IUAC, Delhi where each compute node is built using two Intel Xeon Gold 6240R 2.4 GHz processors with 24 cores each and contains 192 GB RAM, connected via Infiniband interconnect.


\section{Results\label{sec:results}}
The results of the numerical experiments of the three test cases are presented in this section. The two-dimensional isentropic vortex advection has an exact solution and is therefore a suitable problem for verifying the order of accuracy of the asynchronous solver. The three-dimensional Taylor-Green vortex is a transitional flow problem that is widely solved to assess the accuracy of high-order schemes. The breakdown of large-scale structures to smaller scales is sensitive to velocity gradients, and the efficacy of asynchronous solvers in capturing these small-scale structures is investigated. These two test cases are also considered to demonstrate the scalability of asynchronous algorithms by performing strong and weak scaling studies. The third test case is the transitional flow over a NACA0012 airfoil and is considered for evaluating the effect of numerical errors introduced at near wall structures that could perturb the location of the transition to turbulence. Scalability is demonstrated using a strong scaling study to assess the performance under non-uniform workload balancing, commonly associated with simulations of practically relevant problems. For reference, the algorithms used in the numerical experiments are as follows: synchronous algorithm (SA), communication avoiding algorithm (CAA-AS) and synchronization avoiding algorithm (SAA-AS) that use standard schemes in asynchrony, and communication avoiding algorithm (CAA-AT) and synchronization avoiding algorithm (SAA-AT) that implement asynchrony-tolerant schemes at PE boundaries. The numerical schemes used for the algorithms are listed in Table~\ref{tab:lserk_at_schemes}.

\begin{table}[H]
\centering
\begin{tabular}{lllllll}
\hline
\multirow{2}{*}{Case} & \multirow{2}{*}{Algorithm} & \multirow{2}{*}{\begin{tabular}[c]{@{}l@{}}Time\\ stepping\end{tabular}} & \multicolumn{2}{c}{Spatial derivatives} & \multirow{2}{*}{\begin{tabular}[c]{@{}l@{}}Buffer point\\ updates\end{tabular}} & \multirow{2}{*}{\begin{tabular}[c]{@{}l@{}}Overall order \\ of accuracy\end{tabular}} \\
\cmidrule(lr){4-5}
& & & Interior & PE boundary & & \\
\hline
1 & SA & LSERK2 & CD2 & CD2 & - & 2 \\
2 & CAA-AS, SAA-AS & LSERK2 & CD2 & CD2 & - & 1 \\
3 & CAA-AT, SAA-AT & LSERK2 & CD2 & AT2 & AT2 & 2 \\
4 & SA & LSERK4 & CD4 & CD4 & - & 4 \\
5 & CAA-AS, SAA-AS & LSERK4 & CD4 & CD4 & - & 1 \\
6 & CAA-AT, SAA-AT & LSERK4 & CD4 & AT4 & AT4 & 4 \\
\hline
\end{tabular}
\caption{List of schemes in the three algorithms used in numerical experiments, with the expected overall orders of accuracy.}
\label{tab:lserk_at_schemes}
\end{table}

\subsection{Isentropic vortex advection\label{subsec:covo}}
\begin{figure}[h!]
\centering
\includegraphics[height=4.2cm]{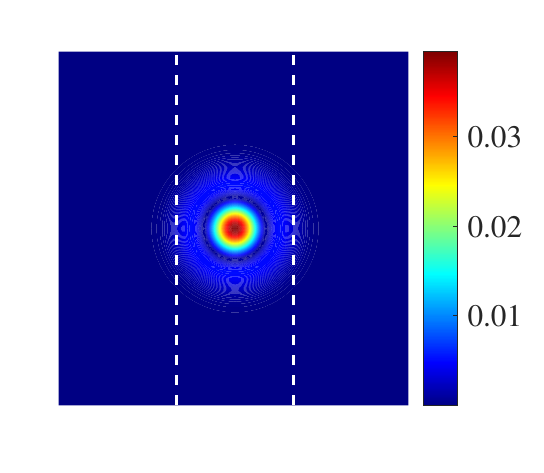}
\begin{picture}(0,0)
\setlength{\unitlength}{1cm}
\put(-4.8,3.4){\bf \small{\textcolor{black}{(a)}}}
\put(-4.55,0.5){\vector(1,0){1}}
\put(-4.55,0.5){\vector(0,0){1}}
\put(-3.55,0.3){\small{x}}
\put(-4.75,1.5){\small{y}}
\put(-4.8,0.25){\small{$-8$}}
\put(-1.45,0.25){\small{$8$}}
\put(-4.5,3.8){\small{$8$}}
\put(-1.15,3.8){\small{$\Omega$}}
\end{picture}
\hspace{-0.4cm}
\includegraphics[height=4.2cm,trim=0cm 0cm 2.5cm 0cm,clip]{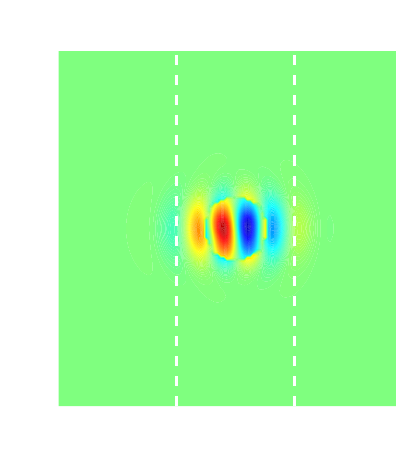}
\begin{picture}(0,0)
\setlength{\unitlength}{1cm}
\put(-3.5,3.4){\bf \small{\textcolor{black}{(b)}}}
\put(-1.75,3.8){SA}
\end{picture}
\hspace{-0.5cm}
\includegraphics[height=4.2cm,trim=0cm 0cm 2.5cm 0cm,clip]{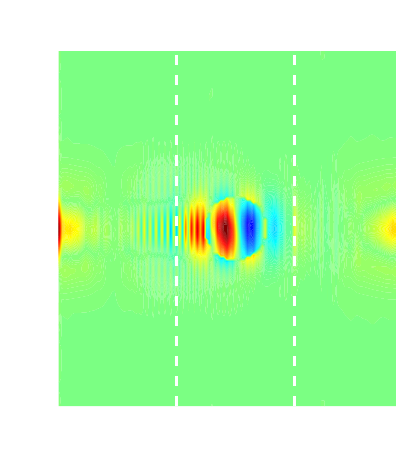}
\begin{picture}(0,0)
\setlength{\unitlength}{1cm}
\put(-3.5,3.4){\bf \small{\textcolor{black}{(c)}}}
\put(-2.2,3.8){CAA-AS}
\end{picture}
\hspace{-0.5cm}
\includegraphics[height=4.2cm]{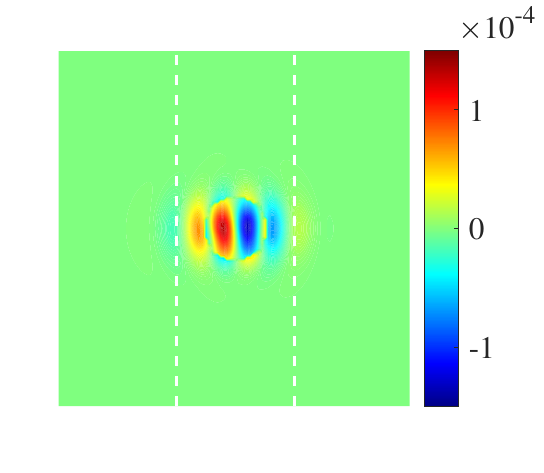}
\begin{picture}(0,0)
\setlength{\unitlength}{1cm}
\put(-4.8,3.4){\bf \small{\textcolor{black}{(d)}}}
\put(-3.5,3.8){CAA-AT}
\put(-0.4,3.3){\scriptsize{\rotatebox{270}{$\sqrt{\frac{1}{N}\sum\limits_{i=1}^{N}(\rho_i-\rho_{\text{ref},i})^2}$}}}
\end{picture}
\caption{(a) Vorticity magnitude contours of initial flow field. Instantaneous contours of errors in vorticity magnitude after one advection cycle for (b) synchronous algorithm (SA), (c) communication avoiding algorithm with standard schemes (CAA-AS), and (d) communication avoiding algorithm with asynchrony-tolerant schemes (CAA-AT), solved with $N_x =N_y = 240$ and $L=8$ using the schemes in cases 4, 5, and 6, respectively, in Table \ref{tab:lserk_at_schemes}. The dashed white lines represent processing element (PE) boundaries.}
\label{fig:vortex_contours}
\end{figure}

\begin{figure}[h!]
\includegraphics[width=7.5cm]{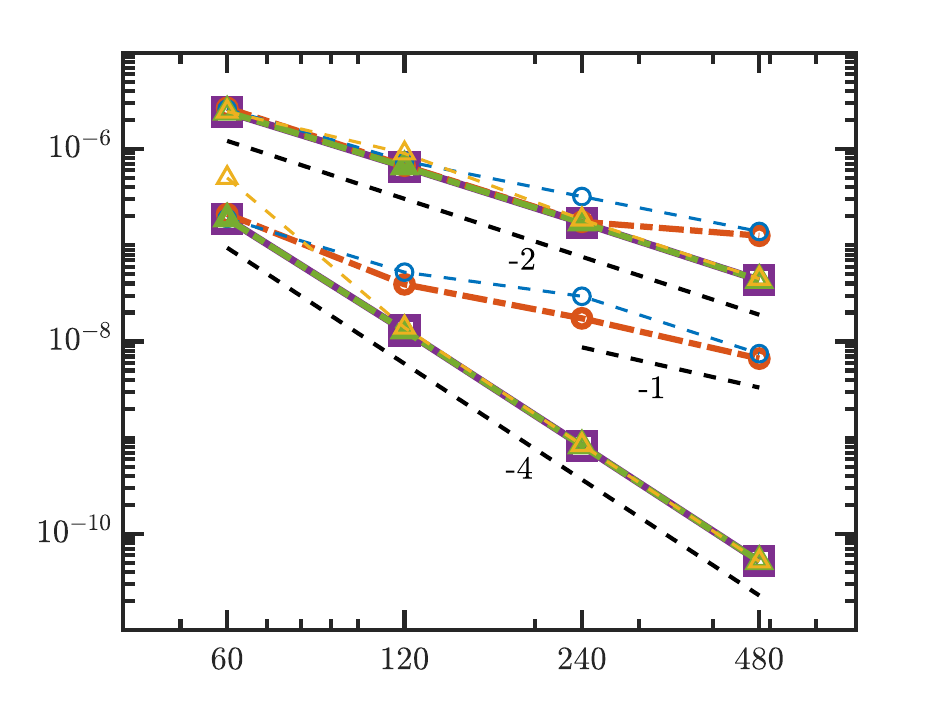} \hspace{0.8cm}
\includegraphics[width=7.5cm]{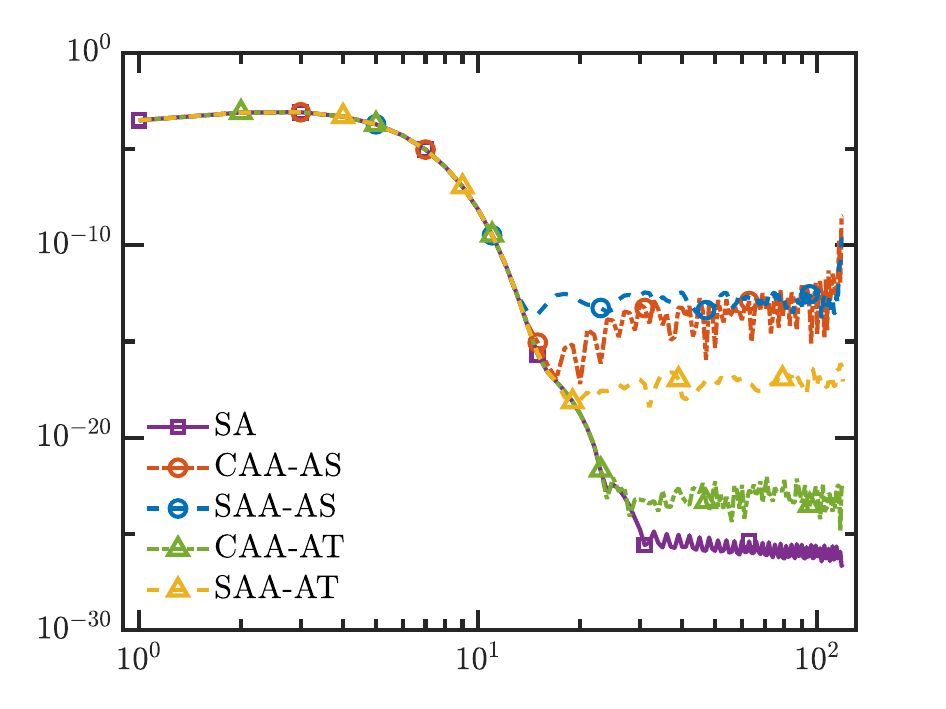}
\begin{picture}(0,0)
    \put(-460,115){\bf \small (a)}
    \put(-220,115){\bf \small (b)}
    \put(-400,-3) {\small Number of grid points $N_x$}
    \put(-460,50){\rotatebox{90}{\small Error $\left< E \right>$}}
    \put(-140,-3) {\small Wave number $\kappa_x$}
    \put(-220,40){\rotatebox{90}{\small Energy $\hat{v}^2(\kappa_x)$}}
\end{picture}
\caption{(a) Order convergence plot for density error for all the cases in Table~\ref{tab:lserk_at_schemes}, and (b) energy spectra of $v$ velocity profile at the $y=0$ line for cases 4, 5, and 6 in Table~\ref{tab:lserk_at_schemes} with $N_x=240$ and $L=8$. The different lines indicate: SA (solid purple), CAA-AS (dash-dotted red), SAA-AS (dashed blue), CAA-AT (dash-dotted green) and SAA-AT (dashed yellow).}
\label{fig:vortex_plots}
\end{figure}

The two-dimensional isentropic vortex advection is an inviscid flow problem with periodic boundary conditions. The computational domain is defined as $-8\leq x \leq 8$ and $-8 \leq y \leq 8$, with the Mach number of the driving flow $M = 0.1$. The vortical flow field is initialized as a Gaussian profile centered at ($x_c,y_c$) \cite{Visbal_JCP_2002}, with the primitive quantities defined as
\begin{equation}
\begin{split}
    u &= 1 - \frac{\beta(y-y_c)}{R^2}\exp\left(\frac{-\Omega^2}{2}\right)\\
    v &= \frac{\beta(x-x_c)}{R^2}\exp\left(\frac{-\Omega^2}{2}\right)\\
    p_{\infty} - p &= \frac{\rho \beta^2}{2R^2}\exp(-\Omega^2)\\
    \Omega^2 &= \frac{(x-x_c)^2+(y-y_c)^2}{R^2},
\end{split}
\label{eq:covo_init}
\end{equation}
where $u$, $v$, $p$ and $R$ denote the $x$ and $y$ velocity components, pressure, and vortex core radius, respectively. The non-dimensional vortex strength parameter $\beta/(U_{\infty}R)$ is chosen as 0.02. Figure~\ref{fig:vortex_contours}(a) plots the contours of the vorticity magnitude at $t=0$ on the domain that is divided into three PEs in the $x$ direction, with PE boundaries denoted by the white dashed lines. The flow is evolved for an entire advection cycle ($t=16$), ensuring that the vortex crosses all three PE boundary interfaces, including the periodic interface, and returns to its initial position. A maximum allowable delay $L=8$ is considered for the asynchronous numerical simulations. Figures~\ref{fig:vortex_contours}(b), (c), and (d) show the contours of the error in the vorticity magnitude (computed against the analytical solution) corresponding to SA, CAA-AS, and CAA-AT, respectively. Time steps of $2\times10^{-3}$, $1\times10^{-3}$ and $1\times10^{-3}$ are used with SA, CAA and SAA, respectively, at $N_x=60$ and are proportionately reduced along with $\Delta x$ to satisfy the stability criterion. The error contours for CAA-AT are very similar to those of SA, whereas larger errors are evident for CAA-AS, particularly near the PE boundaries. To quantify the overall accuracy, the mean error $\langle E \rangle$ in density, which is obtained by averaging the RMS error over all grid points, across multiple independent simulations to account for the randomness in the delays, at various grid resolutions is considered. From the numerical experiments, the second- (schemes in cases 1 and 3 in Table~\ref{tab:lserk_at_schemes}) and fourth-order (schemes in cases 4 and 6 in Table~\ref{tab:lserk_at_schemes}) of CAA-AT and SAA-AT are verified, as shown in the error convergence plot in Fig.~\ref{fig:vortex_plots}(a). However, the errors for CAA-AS and SAA-AS (cases 2 and 5 in Table~\ref{tab:lserk_at_schemes}) collapse to the first-order, thereby confirming the earlier predictions \cite{Donzis_JCP_2014}.

\begin{figure}[h!]
\includegraphics[width=7.5cm]{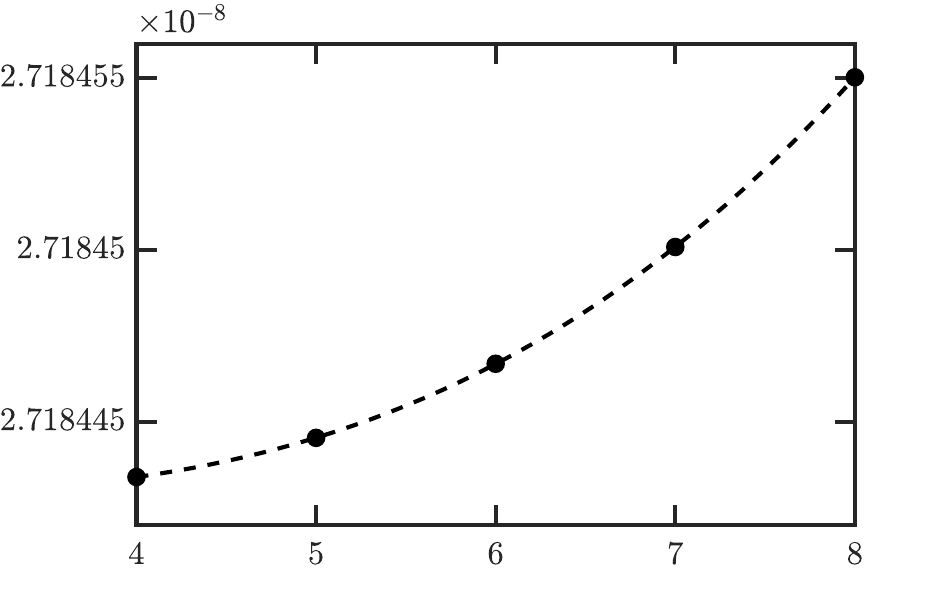} \hspace{0.8cm}
\includegraphics[width=7.5cm]{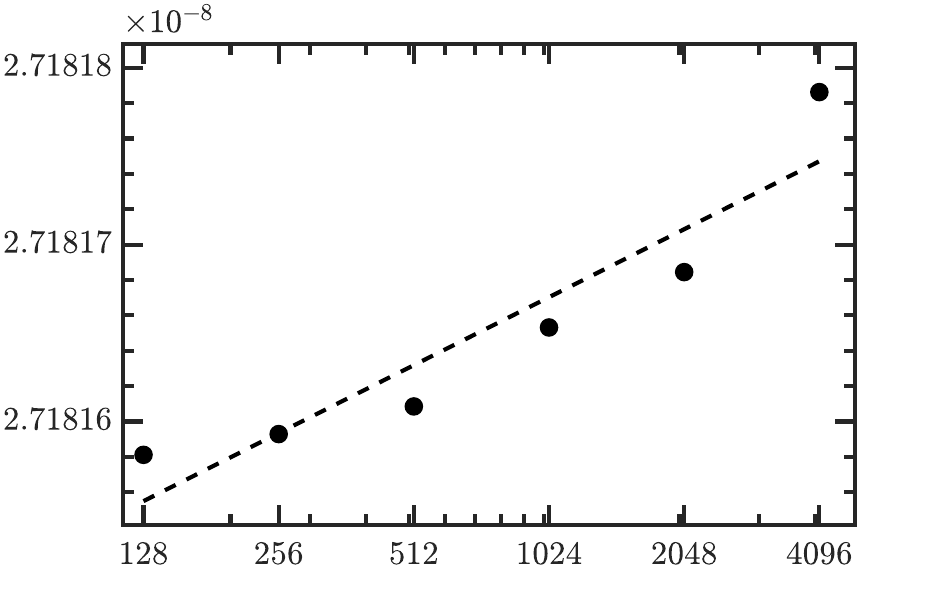}
\begin{picture}(0,0)
    \put(-460,100){\bf \small (a)}
    \put(-220,100){\bf \small (b)}
    \put(-350,-3) {\small $L$}
    \put(-470,50){\rotatebox{90}{\small Error $\left< E \right>$}}
    \put(-110,-3) {\small $P$}
    \put(-230,50){\rotatebox{90}{\small Error $\left< E \right>$}}
\end{picture}
\caption{Plots of error in $u$-velocity after $t=4$ for the isentropic vortex advection solved on a $2048^2$ grid against (a) maximum allowable delay $L$ on 1024 PEs and (b) number of PEs $P$ with $L=4$ for CAA-AT corresponding to case 3 in Table~\ref{tab:lserk_at_schemes}. Dashed lines in (a) and (b) represent best fit cubic polynomial in $L$, and reference linear polynomial respectively.}
\label{fig:vortex_errorvsL_P}
\end{figure}

The accuracy of the numerical solutions is further examined by plotting the energy spectra since the errors incurred due to communication delays at the PE boundaries are localized in space and are expected to affect the solution at high wavenumbers. In Fig.~\ref{fig:vortex_plots}(b), plotted against the wave number $\kappa_x$ are the energy spectra ($\hat{v}^{2}_{}(\kappa_x)$) of $v$-velocity along the $y=0$ line obtained from the numerical solutions computed using the cases 4, 5 and 6 in Table~\ref{tab:lserk_at_schemes}. In general, the physical spectra remain approximately constant at lower wavenumbers and begin decaying as $\kappa_x$ increases. For the baseline SA, the physical spectrum is captured until $\kappa_x \approx 30$ and the energy accumulated at higher wavenumbers is due to round-off errors. In comparison, the spectra of CAA-AS and SAA-AS start deviating from those of SA at a smaller wavenumber $\kappa_x \approx 15$ and contain similar levels of energy, which are more than 10 orders of magnitude higher than those of SA. These deviations are a consequence of high-frequency numerical errors in the solution due to asynchrony at the PE boundaries. Meanwhile, the spectrum of SAA-AT deviates from that of SA at $\kappa_x \approx 20$ with slightly less energy contributions at higher wavenumbers. CAA-AT resolves the solution accurately until $\kappa_x \approx 25$ and contains similar higher wavenumber behavior to SA. The difference between the spectra of SAA-AT and CAA-AT can be attributed to the randomness of the communication delays in SAA-AT, which agrees with the results of previous studies \cite{Komal_JCP_2020}.

Unlike the synchronous algorithm, the error in the asynchronous algorithms depends on simulation parameters such as the maximum allowable delay ($L$) and number of PEs ($P$), in addition to the resolution. The error scaling relation $\langle \bar{E} \rangle \propto (P/N)\Delta x^{q}\sum_{m=1}^{C}\gamma_{m}\bar{\tilde{k}}^{m}$ was derived in a previous study based on theoretical analyses \cite{Aditya_JCP_2017}, where $\bar{\tilde{k}}$ and $\gamma_m$ denote the mean delay and the coefficient of its $m$th moment, respectively. To validate this error scaling, simulations are performed on a $2048^2$ grid with two-dimensional domain decomposition, till $t=4$ using Case 3 in Table~\ref{tab:lserk_at_schemes}, considering different values of $P$ and $L$. For CAA-AT, the delays are periodic and follow a uniform distribution.
In the case considered here, the number of delayed time levels in the AT stencil is $C=3$ and the error therefore scales as a cubic polynomial in $L$ and linearly with $P$. Figure~\ref{fig:vortex_errorvsL_P}(a) plots the variation of $u$-velocity error against $L$ with a fixed value of $P=1024$, while Fig.~\ref{fig:vortex_errorvsL_P}(b) plots the variation of error against $P$ with a fixed value of $L=4$. A cubic polynomial is fit to the plot in Fig.~\ref{fig:vortex_errorvsL_P}(a) while a line is fit to the plot in Fig.~\ref{fig:vortex_errorvsL_P}(b), agreeing with the predictions.

\begin{figure*}[h]
\centering
\includegraphics[width=6cm]{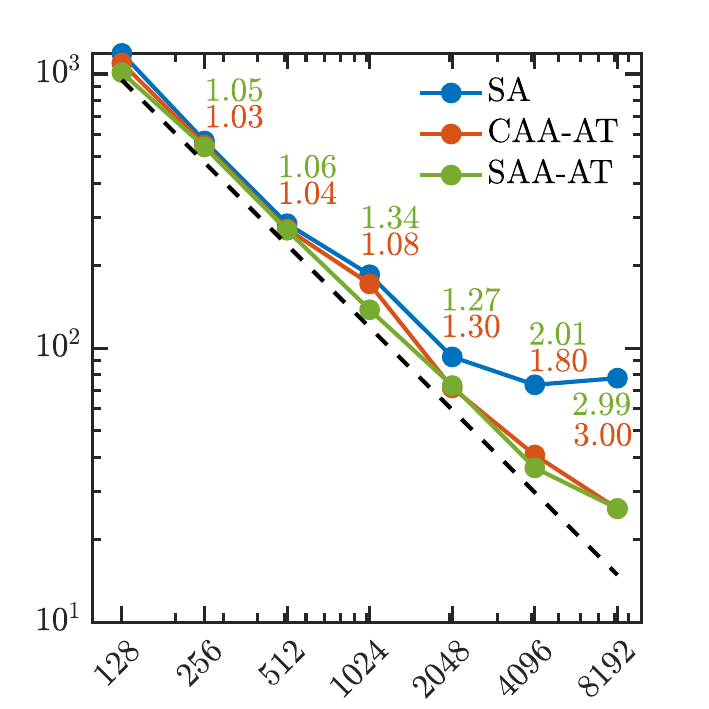} \hspace{1cm}
\includegraphics[width=6cm]{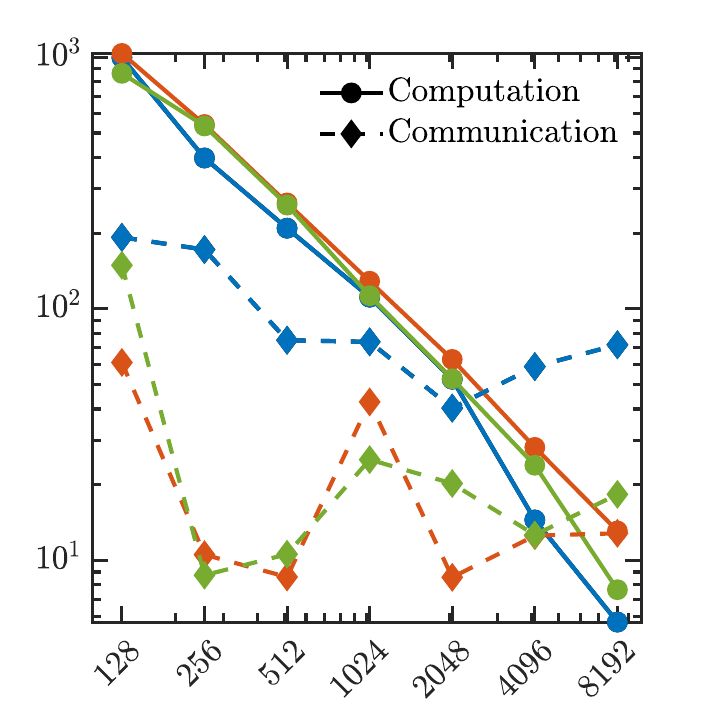}
\begin{picture}(0,0)
    \put(-375,140){\bf \small (a)}
    \put(-170,140){\bf \small (b)}
    \put(-325,-5){\small Number of PEs}
    \put(-120,-5){\small Number of PEs}
    \put(-385,40){\rotatebox{90}{\small Total execution time (s)}}
    \put(-180,80){\rotatebox{90}{\small Time (s)}}
\end{picture}
\caption{Strong scaling results of isentropic vortex advection on a $16384^2$ grid with cases 4 and 6 in Table~\ref{tab:lserk_at_schemes}. (a) Total execution time, and (b) split between time taken for computation (solid lines) and communication (dashed lines) against number of PEs, for SA (blue), CAA-AT (red) and SAA-AT (green). $L=10$ for CAA-AT and SAA-AT.}
\label{fig:covo_strong}
\end{figure*}

The computational performance of the asynchronous algorithms is evaluated using a strong scaling study, in which the overall problem size is kept fixed and the number of PEs is increased. Here, a two-dimensional grid of size $16384\times 16384$ is decomposed into subdomains along both $x$ and $y$ directions ensuring uniform workload distribution. The execution times along with the contribution from computations and communications are recorded by solving for 100 time steps using the cases 4 and 6 in Table~\ref{tab:lserk_at_schemes}, with a maximum allowable delay of 10. This procedure is repeated over 10 independent trials and averaged to account for performance variation. Figure~\ref{fig:covo_strong}(a) plots the total execution time against the number of CPU cores. The speed-ups of CAA-AT and SAA-AT relative to SA at each scale are shown in the figure. It appears that the total execution time of SA deviates early from ideal scaling (denoted by the black dashed line with slope -1) and plateaus at 8192 cores, while both CAA-AT and SAA-AT continue to remain close to the ideal scaling line. In fact, both asynchronous algorithms achieve a speed-up of up to $3\times$ compared with the baseline SA-based solver at the extreme scale. This can be explained through the distribution of computation and communication times for different numbers of cores, as shown in Fig.~\ref{fig:covo_strong}(b). For all three cases, the computation time drops linearly with an increase in the PEs. Note that CAA-AT and SAA-AT incur a small additional overhead in computation compared to SA. Meanwhile, the communication cost of SA is significantly greater than that of the two asynchronous algorithms. Furthermore, the communication time overwhelms the computation time at 2048 cores for SA, where the scaling of the total execution time plateaus in part (a) of the figure. These plots demonstrate the significantly reduced communication costs associated with the asynchronous algorithms, resulting in improved scalability.

\subsection{Taylor-Green vortex\label{subsec:tgv}}

The Taylor-Green vortex (TGV) is a widely considered benchmark problem for assessing the numerical accuracy of high-order schemes. This test case contains key physical phenomena such as vortex dynamics, transition to turbulence and turbulent dissipation, facilitated by an initial periodic flow defined in Eq.~\ref{eq:tgv_ic} based on \cite{Diosady_report_2015,Debonis_AIAA_2013}.
\begin{equation}
\begin{split}
    u &= U_0\sin(x/L_{0})\cos(y/L_{0})\cos(z/L_{0})\\
    v &= U_0\cos(x/L_{0})\sin(y/L_{0})\cos(z/L_{0})\\
    w & = 0\\
    p &= \rho_0U_0^2\left[\frac{1}{\gamma M_0^2} + \frac{1}{16}\left(\cos(2x/L_{0})+\cos(2y/L_{0})\right)\left(\cos(2z/L_{0})+2\right)\right]
\end{split}
\label{eq:tgv_ic}
\end{equation}
The domain has dimensions $L_x \times L_y \times L_z = 2\pi \times 2\pi \times 2\pi$ with periodic boundary conditions. The Mach number and Reynolds number are taken to be $M_0 = 0.1$ and $Re = \rho_{0}U_{0}L_{0}/\mu_{0} = 1600$, to validate with the studies carried out in the literature \cite{Diosady_report_2015,Debonis_AIAA_2013}. A $512^3$ grid is divided uniformly in all three directions into 512 PEs and is solved using the SA, CAA-AS, and CAA-AT algorithms with schemes in cases 4, 5, and 6 from Table~\ref{tab:lserk_at_schemes}, respectively. Tenth-order implicit filtering with a filter coefficient $\alpha_{f}=0.499$ is used to prevent numerical oscillations and ensure the stability of the solver. The maximum allowable delays ($L$) of values $2$ and $5$ are considered for the CAA-AS and CAA-AT, respectively. Note that because AT schemes provide better stability relative to standard schemes with delays (AS), a greater $L$ can be imposed. The simulations are performed in the following manner: the standard synchronous algorithm (SA) is used over the time interval $0\leq t/t_c\leq 8$ with a time step $1.0\times 10^{-3}$, where $t_c=L_{0}/U_{0}$ is the characteristic convective time. The checkpoint at $t/t_c=8$ is used to initialize the flow, which is solved with SA, CAA-AS, and CAA-AT using a smaller time step $1.0\times 10^{-4}$ due to the stability constraints associated with the asynchronous computing approach, particularly when standard schemes are employed with delays. The time interval to perform the comparison study is chosen to be $8\leq t/t_c\leq 12$, where the enstrophy reaches its peak value due to the vortex interactions and the transition to turbulence.

\begin{figure}[h!]
\centering
\includegraphics[width=8cm]{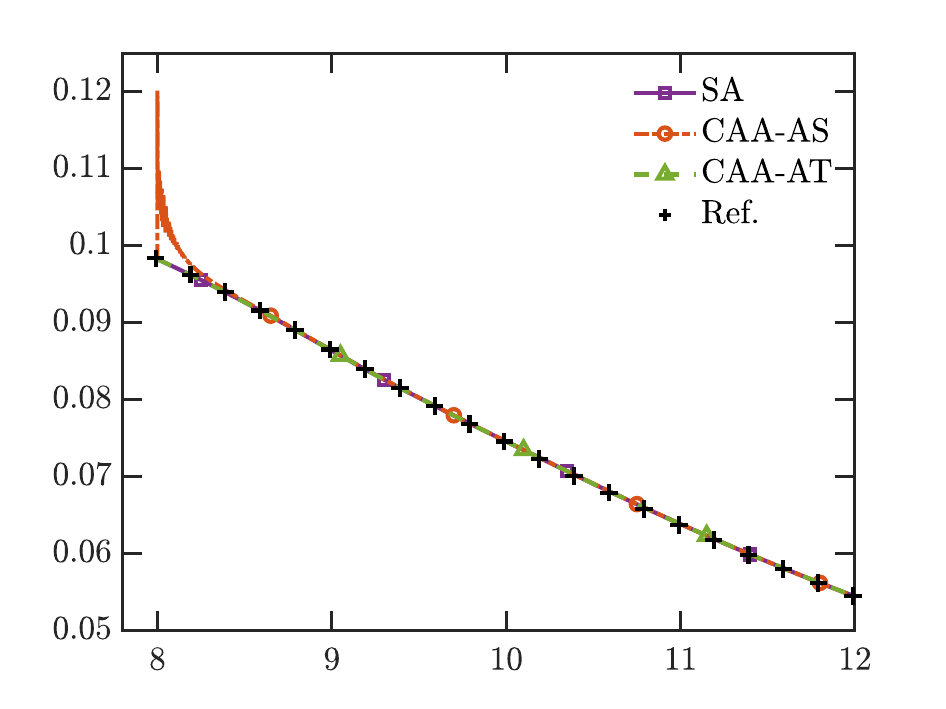}
\includegraphics[width=8cm]{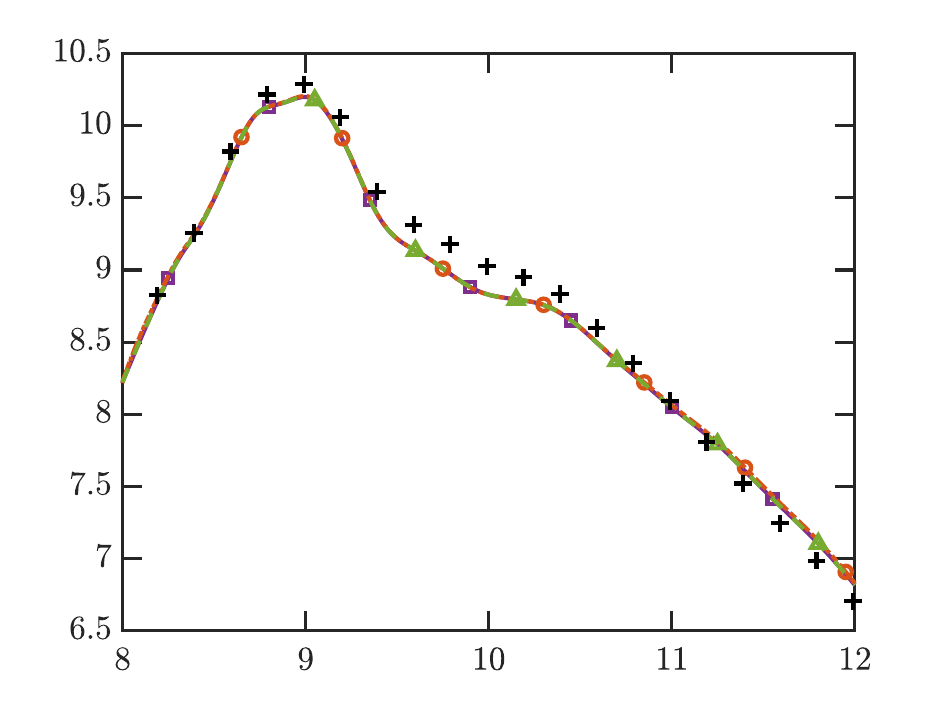}
\begin{picture}(0,0)
\put(-120,-0){\small $t/t_c$}
\put(-345,-0){\small $t/t_c$}
\put(-460,55){\rotatebox{90}{\small Kinetic energy}}
\put(-225,65){\rotatebox{90}{\small Enstrophy}}
\put(-465,140){\bf \small (a)}
\put(-230,140){\bf \small (b)}
\put(-277,117){\small \cite{vanRees_JCP_2011}}
\end{picture}
\caption{Time evolution of volume-averaged (a) kinetic energy and (b) enstrophy on $512^3$ grid for the Taylor-Green vortex using cases 4, 5, and 6 in Table~\ref{tab:lserk_at_schemes}. Maximum allowable delay $L$ is $2$ and $5$ for CAA-AS and CAA-AT respectively.}
\label{fig:tgv_evolution}
\end{figure}

\begin{figure}[h!]
\centering
\hspace{-1cm}
\includegraphics[width=4cm,trim=20cm 2.5cm 20cm 19.5cm,clip]{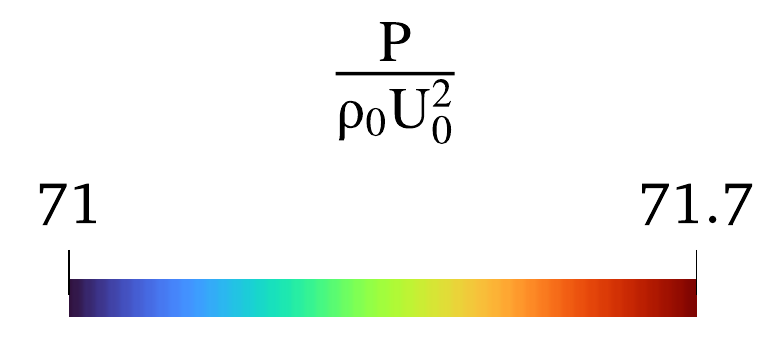}\\
\hspace{-0.9cm}
\includegraphics[height=5cm,trim=14cm 0cm 13cm 2cm,clip]{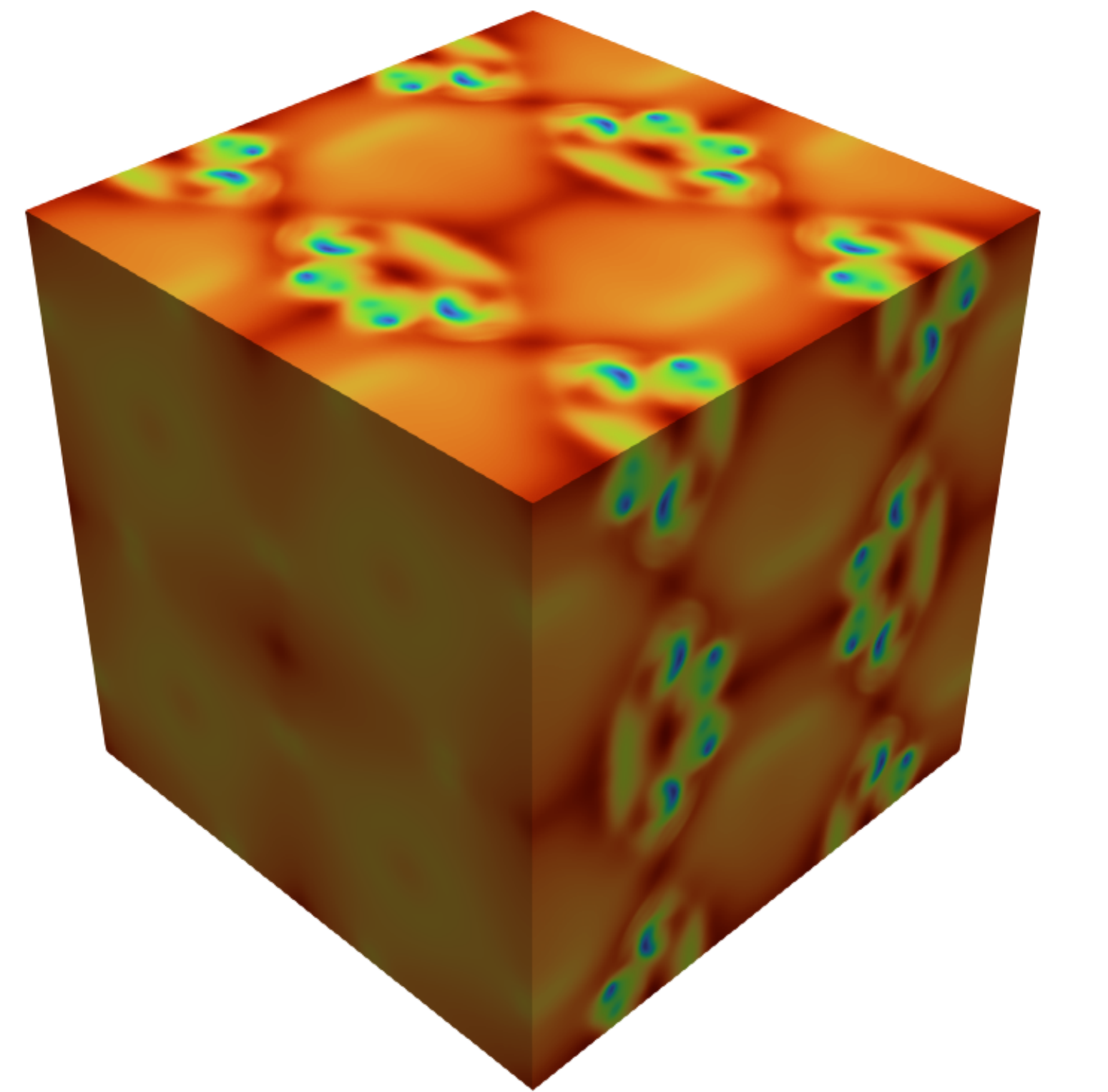}\hspace{0.cm}
\includegraphics[height=5cm,trim=14cm 0cm 13cm 2cm,clip]{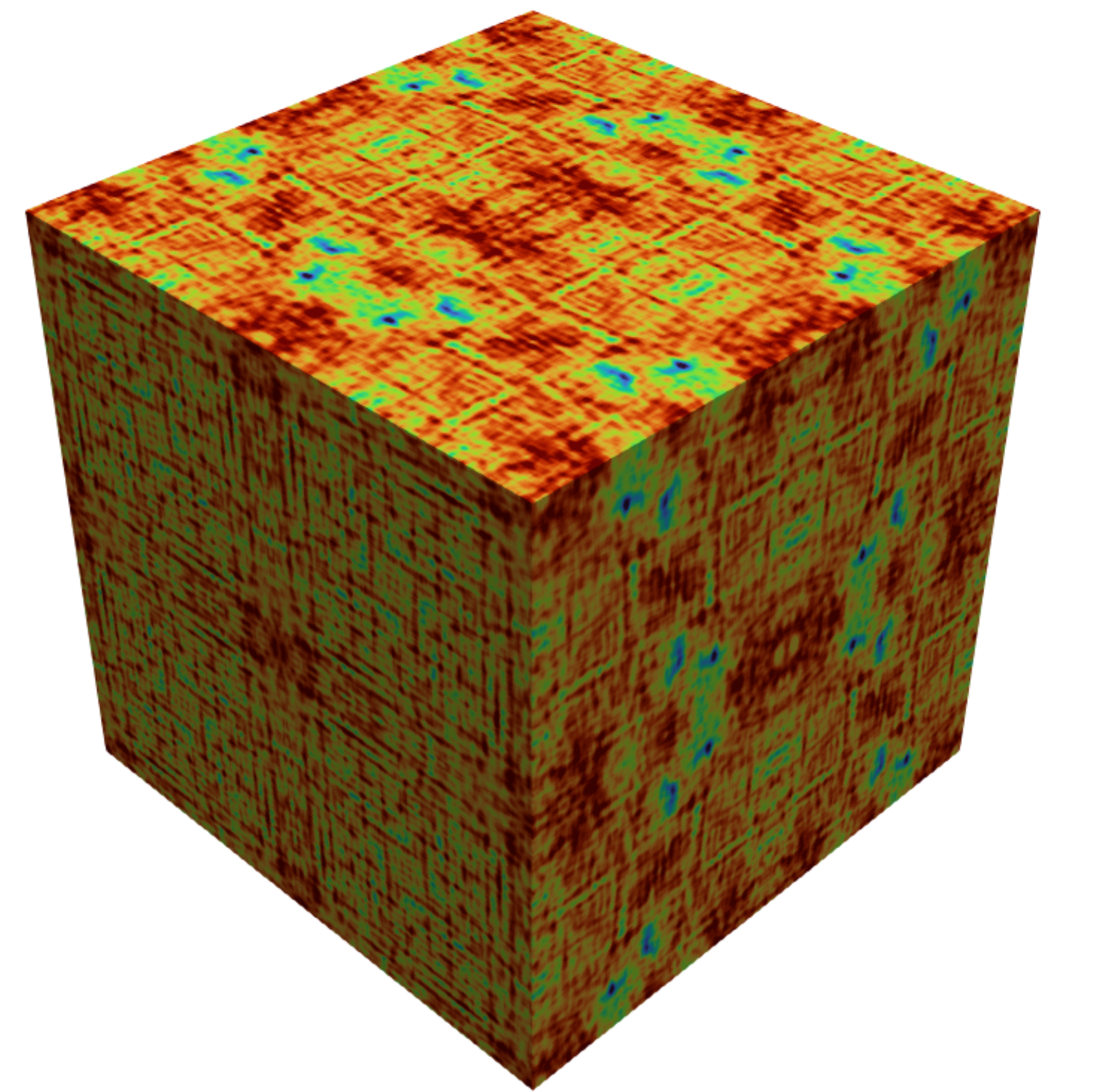}\hspace{0.1cm}
\includegraphics[height=5cm,trim=14cm 0cm 13cm 2cm,clip]{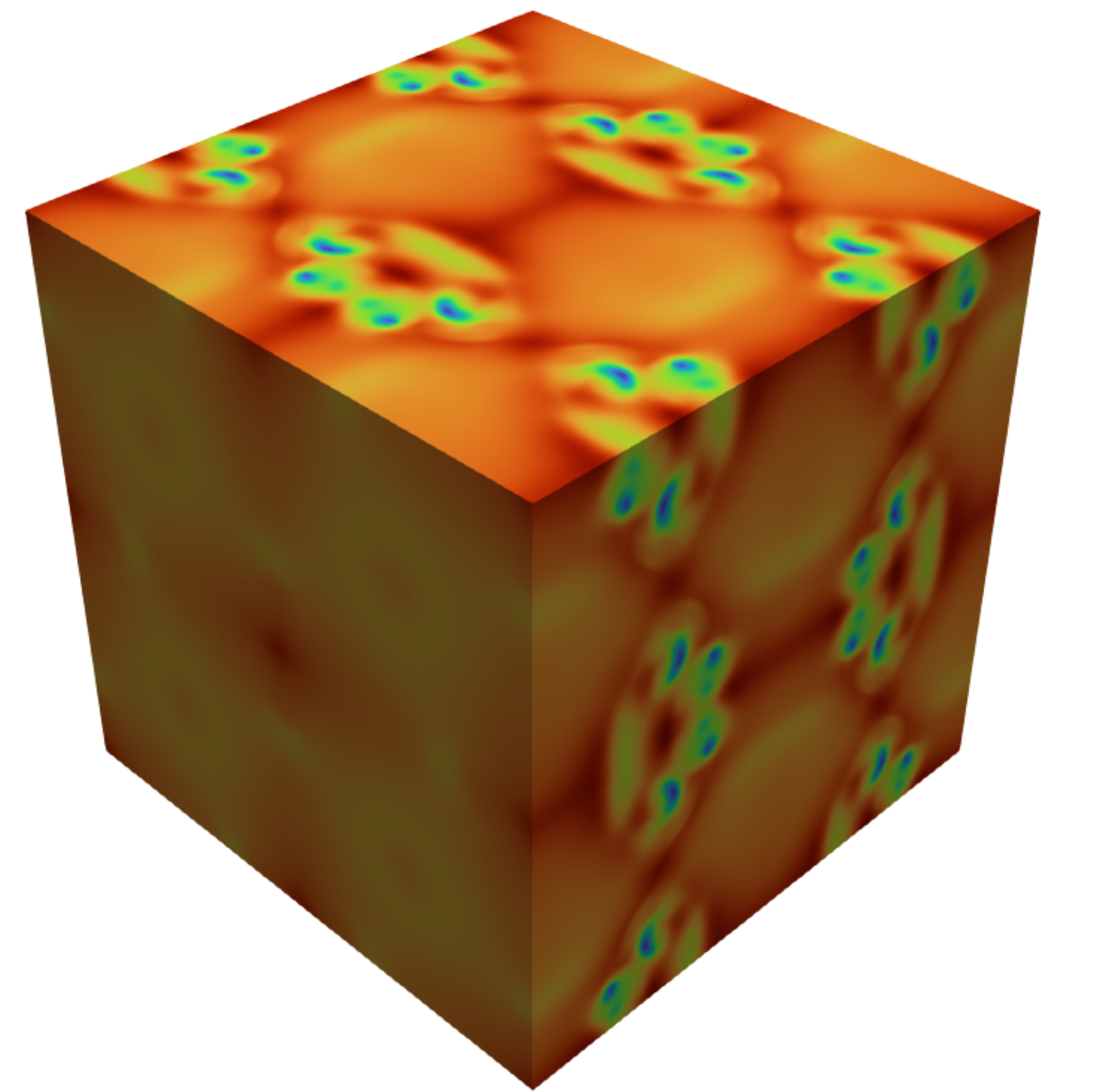}
\\
\includegraphics[height=8cm,trim=0cm 0cm 2.2cm 0cm,clip]{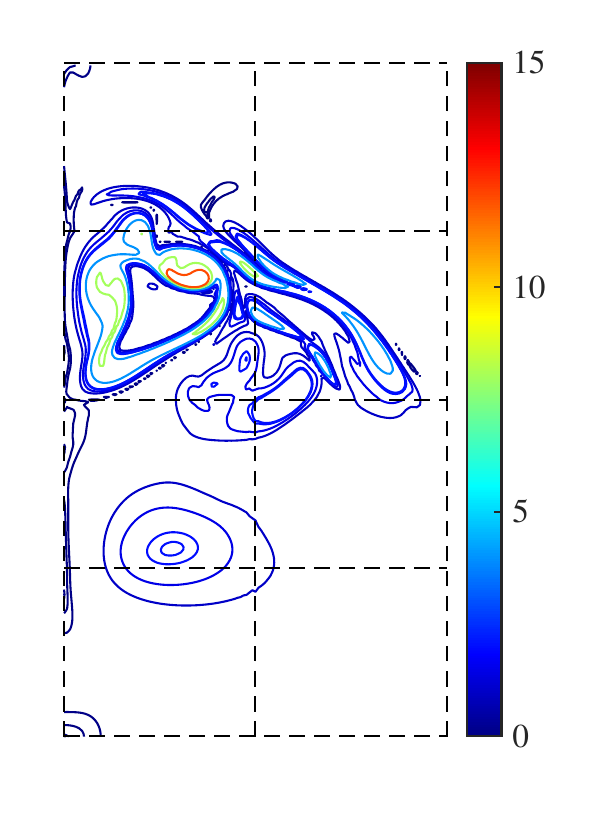}\hspace{0.5cm}
\includegraphics[height=8cm,trim=0cm 0cm 2.2cm 0cm,clip]{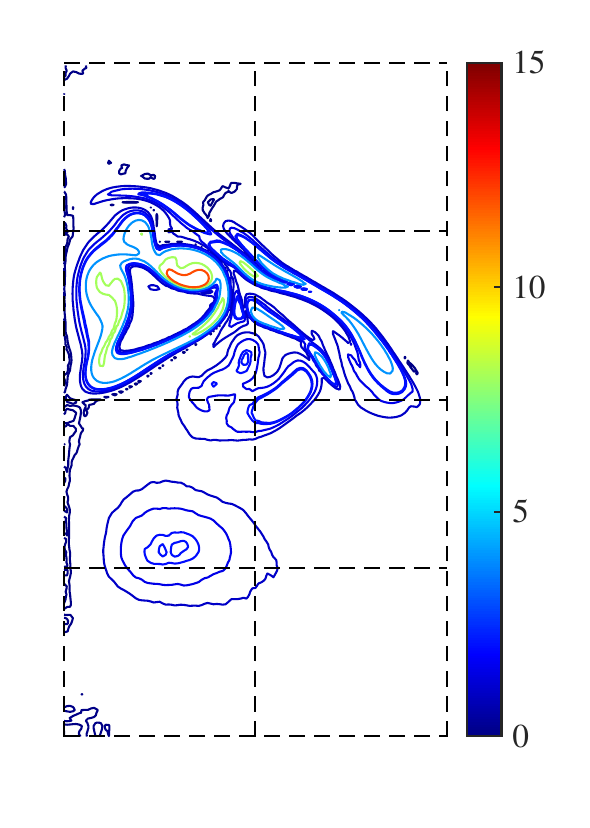}\hspace{0.75cm}
\includegraphics[height=8cm]{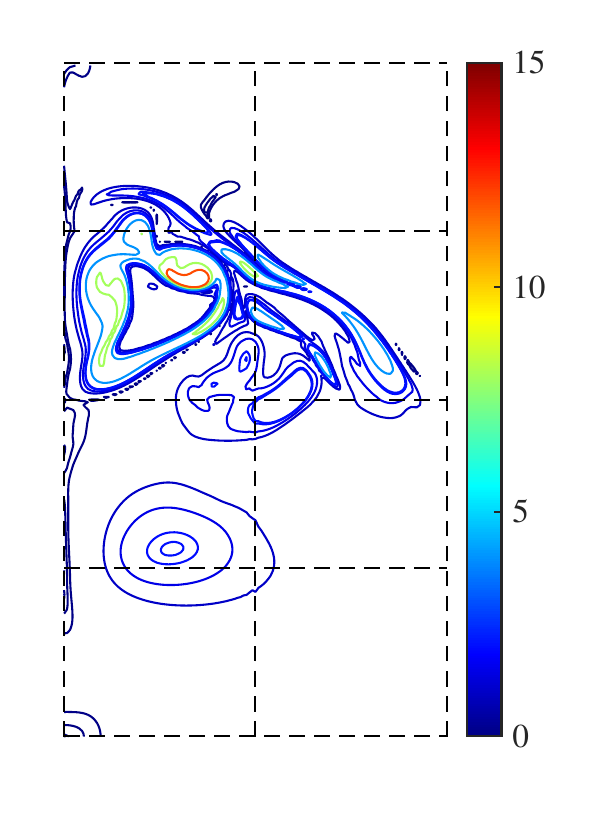}
\begin{picture}(0,0)
\put(-400,215){\large SA}
\put(-270,215){\large CAA-AS}
\put(-120,215){\large CAA-AT}
\put(-450,380){\bf \small (a)}
\put(-450,220){\bf \small (b)}
\put(-20,115){\rotatebox{0}{$\dfrac{\Omega L_0}{U_0}$}}
\end{picture}
\caption{(a) Global contours of pressure, and (b) magnified contours of vorticity magnitude around a vortical structure at the $x=0$ plane, at $t/t_c=9$ with SA, CAA-AS and CAA-AT corresponding to cases 4, 5 and 6 in Table~\ref{tab:lserk_at_schemes}. Maximum allowable delay $L$ is $2$ and $5$ for CAA-AS and CAA-AT respectively. The dashed black lines represent processing element (PE) boundaries.}
\label{fig:tgv_contours}
\end{figure}

The volume averaged kinetic energy, $K = 1/(2\rho_{0}\mathcal{V})\int_{\mathcal{V}}\rho u_{i}u_{i}d\mathcal{V}$, and enstrophy, $\epsilon = 1/(2\rho_{0}\mathcal{V})\int_{\mathcal{V}}\rho\omega_{i}\omega_{i}d\mathcal{V}$, are computed over this time interval, where $\omega_i$ is vorticity and $\mathcal{V}$ is the volume of the domain. The evolution of $K$ and $\epsilon$ are plotted in Fig.~\ref{fig:tgv_evolution}(a) and (b), respectively, for SA, CAA-AS, and CAA-AT, along with the results from pseudo-spectral method-based solutions \cite{vanRees_JCP_2011} for reference. CAA-AS is largely in agreement with SA, except for minor deviations in the evolution of enstrophy and a sudden initial spike in kinetic energy. On the other hand, CAA-AT is in excellent agreement with SA throughout the simulation period, and both are similar to the reference results.

We recall again that errors due to communication delays are introduced in a localized manner near PE boundaries; hence, it is important to inspect the flow structures with particular emphasis on their behavior near PE boundaries. The solution at $t/t_c=9$ is considered for the analysis because, as shown in Fig.~\ref{fig:tgv_evolution}(b), the enstrophy at this instant is close to its maximum. The pressure contours in Fig.~\ref{fig:tgv_contours}(a) highlight the spurious pressure waves throughout the domain in CAA-AS, which can be attributed to the communication delays. The pressure fluctuations in the global flow field are expected to affect the velocity fields and their gradients. However, the contours of CAA-AT are accurately represented without any visible spurious oscillations and are similar to those of SA. Figure~\ref{fig:tgv_contours}(b) plots the magnified contours of the vorticity magnitude around a key vortical structure at the $x=0$ plane that has been studied extensively in the literature \cite{vanRees_JCP_2011}. A close inspection of the contour lines near the PE boundaries (denoted by the dashed black line at the left boundary) shows distortion of the vortical structures in the case of CAA-AS. The contours of CAA-AT, on the other hand, are in good agreement with those of SA, highlighting the effect of AT schemes in reducing numerical errors due to communication delays, especially near the PE boundaries.

\begin{figure}[]
\centering
\includegraphics[width=6cm]{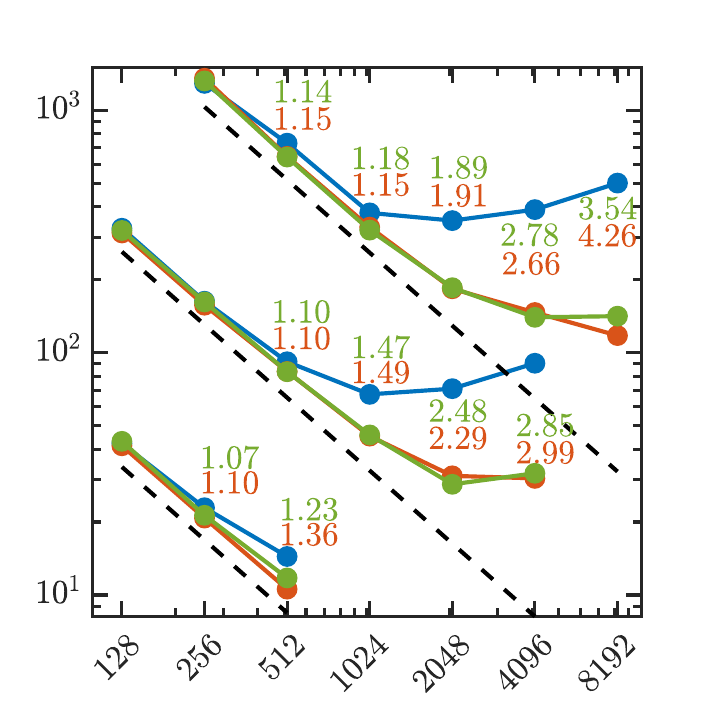}\hspace{1cm}
\includegraphics[width=6cm]{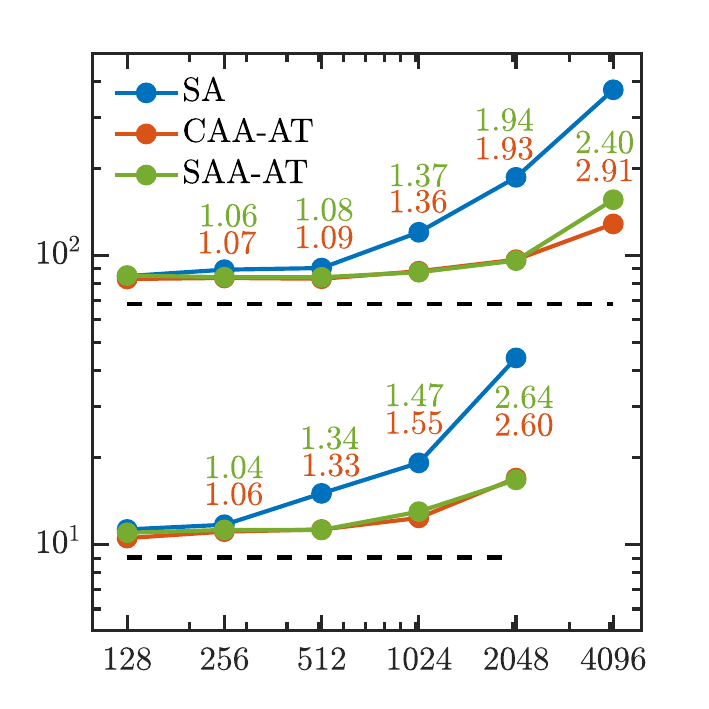}
\begin{picture}(0,0)
    \put(-385,150){\bf \small (a)}
    \put(-180,150){\bf \small (b)}
    \put(-325,-5){\small Number of PEs}
    \put(-120,-5){\small Number of PEs}
    \put(-385,40){\rotatebox{90}{\small Total execution time (s)}}
    \put(-180,40){\rotatebox{90}{\small Total execution time (s)}}
\end{picture}
\caption{(a) Strong scaling results for resolutions of $256^3$, $512^3$ and $1024^3$, and (b) weak scaling results for resolutions of $32^3$ and $64^3$ per PE, for the Taylor-Green vortex, with SA, CAA-AT and SAA-AT (see cases 4 and 6 in Table~\ref{tab:lserk_at_schemes}). Maximum allowable delay $L=10$ for CAA-AT and SAA-AT.}
\label{fig:tgv_scaling}
\end{figure}

The scalability of the asynchronous algorithms in the case of a three-dimensional domain decomposition is demonstrated using strong and weak scaling experiments with the TGV case. In the strong scaling study, grids resolutions of $256^3$, $512^3$ and $1024^3$ are decomposed into subdomains along all three directions and solved on different core counts. In the weak scaling study, the number of grid points per PE is kept fixed, whereas the number of PEs is increased. Ideally, the overall problem is expected to be solved at a constant time irrespective of the number of PEs. Two configurations, $32^3$ and $64^3$ grid points per PE, are considered in this study. Figure~\ref{fig:tgv_scaling} plots the total execution time recorded for 100 time steps against the number of PEs for strong and weak scaling. A maximum allowable delay $L=10$ is considered for CAA-AT and SAA-AT in both studies. The experiments are performed over 10 independent trials, and the average execution time is reported to account for the variation in performance. In strong scaling (see Fig.~\ref{fig:tgv_scaling}(a)), the standard SA begins to deviate from ideal scaling (denoted by dashed black lines of $-1$ slope) and plateaus out at 2048 cores, whereas CAA-AT and SAA-AT lie closer to ideal scaling at the same scale. Moreover, the execution time for SA increases at extreme scales due to the overwhelming communication/synchronization overheads, as discussed in the scaling analysis in Sec.~\ref{subsec:covo}. Speed-ups of $4.26\times$ and $3.54\times$ relative to SA are observed for CAA-AT and SAA-AT, respectively, at the extreme scale of 8192 cores. Similarly, in the weak scaling plot in Fig.~\ref{fig:tgv_scaling}(b), SA starts deviating significantly from the ideal line at scales of 512 and 1024 cores for $32^3$ and $64^3$ grid points per core, respectively, whereas the asynchronous algorithms again scale better than SA for both the configurations. Here, CAA-AT and SAA-AT achieve maximum speed-up of $2.91\times$ and $2.64\times$, respectively, relative to SA. In both strong and weak scaling, CAA-AT achieves slightly better speed-ups compared to SAA-AT at the extreme scales, which is consistent with the trend observed in previous studies \cite{Shubham_JCP_2023}. 

\subsection{Flow over NACA0012 airfoil\label{subsec:naca_transition}}

\begin{figure}
\centering
\includegraphics[width=8cm,trim=0cm 2cm 0cm 0cm,clip]{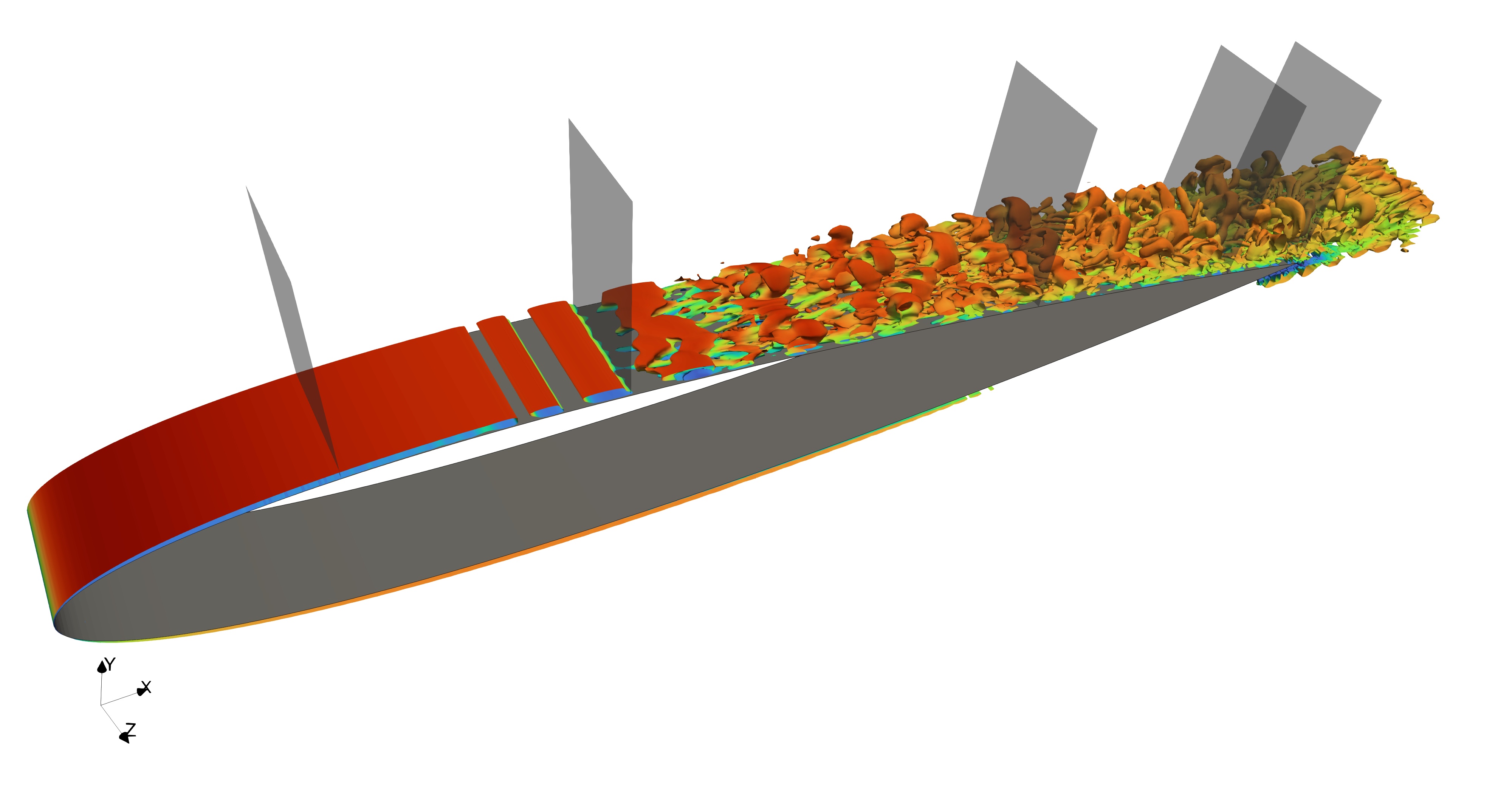}
\includegraphics[width=8cm,trim=0cm 2cm 0cm 0cm,clip]{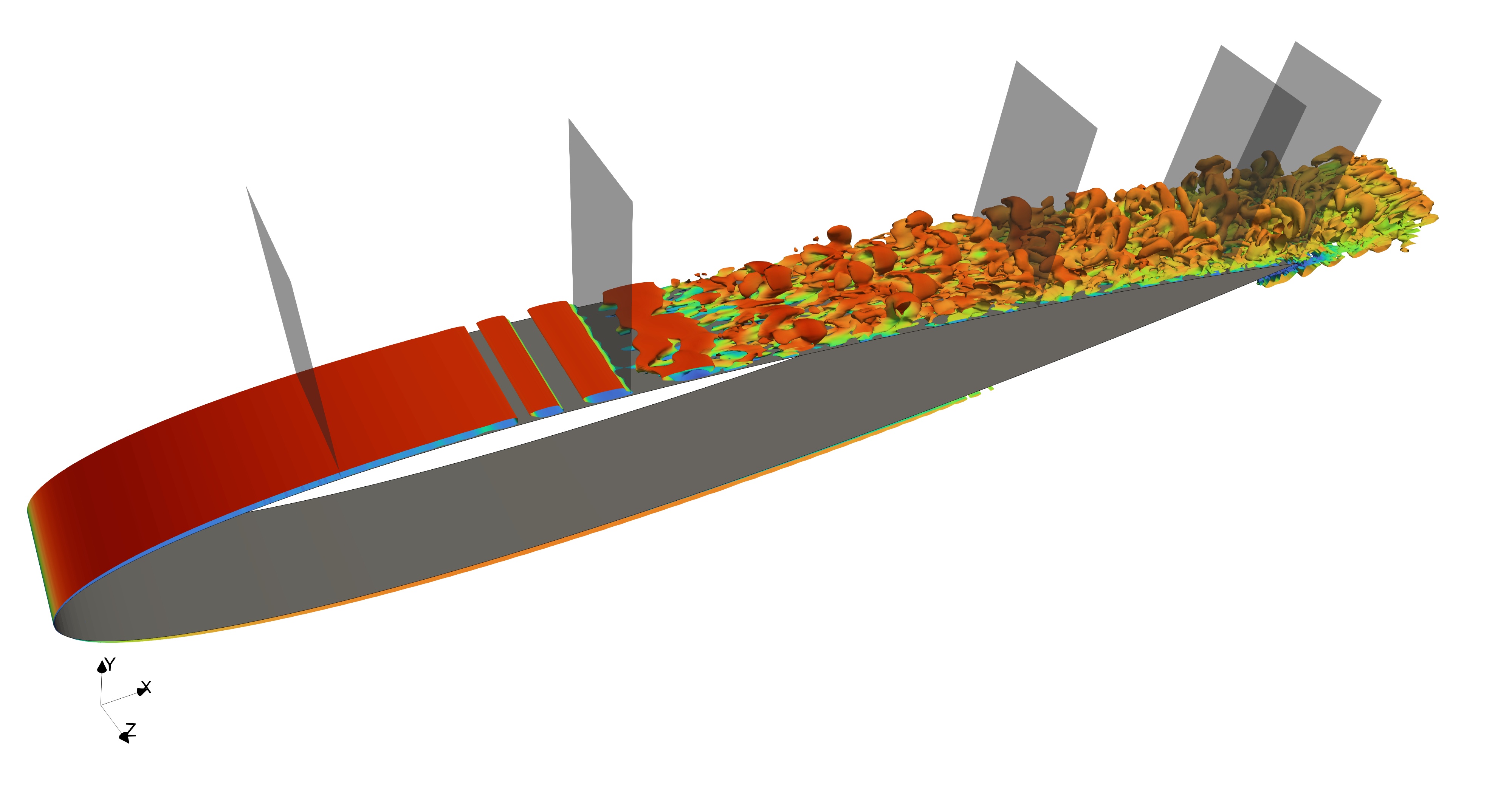}
\includegraphics[width=3cm]{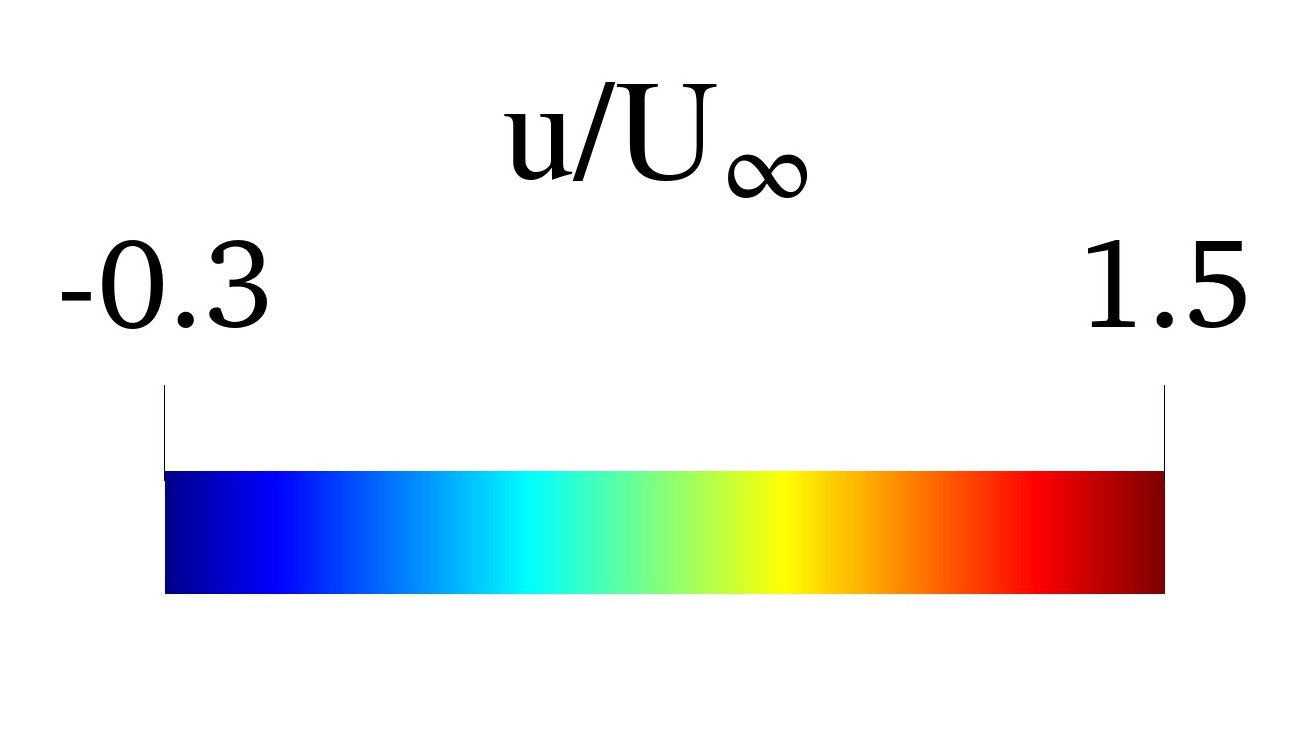}
\begin{picture}(0,0)
\put(-260,140){\bf \small (a)}
\put(-30,140){\bf \small (b)}
\put(-200,140){\color{black}{\vector(1,-2){10}}}
\put(-235,144){KH rollers}
\put(-158,88){\color{black}{\vector(-1,2){15}}}
\put(-170,80){Secondary instabilities}
\put(88,152){\color{black}{\vector(0,-1){21}}}
\put(88,152){\color{black}{\vector(1,-4){6.5}}}
\put(40,155){Hairpin vortices}
\end{picture}
\caption{Instantaneous iso-surfaces of $Q=10$ for the flow around NACA0012 airfoil at AoA $4\degree$ with (a) SA and (b) CAA-AT (cases 4 and 6, respectively, in Table~\ref{tab:lserk_at_schemes}), colored by contours of $u/U_{\infty}$. Maximum allowable delay $L=1$ for CAA-AT.}
\label{fig:airfoil_q_iso}
\end{figure}

\begin{figure}
\centering
\includegraphics[width=8cm,trim=0.1cm 0.1cm 0.1cm 0.1cm,clip]{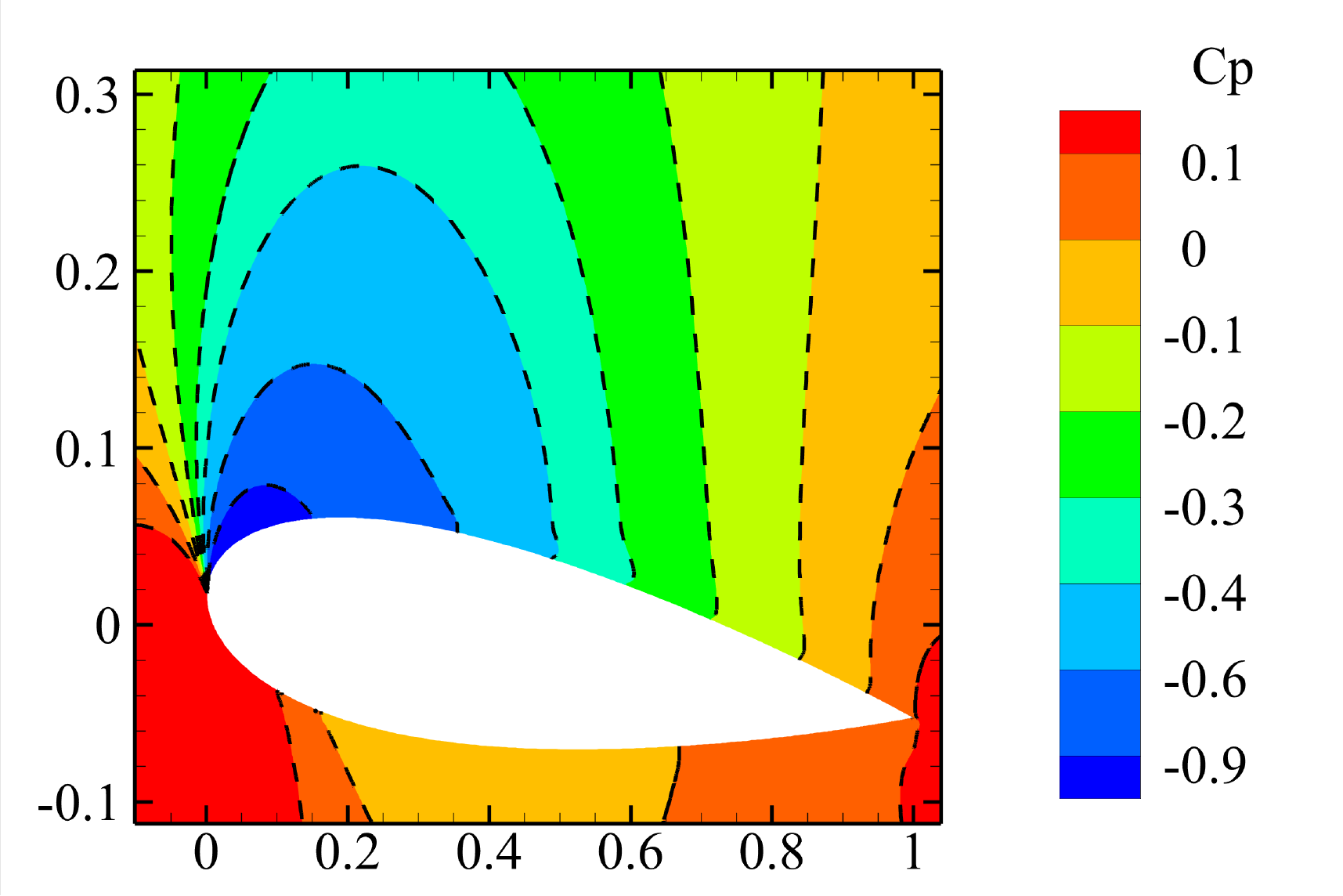}
\includegraphics[width=8cm,trim=0.1cm 0.1cm 0.1cm 0.1cm,clip]{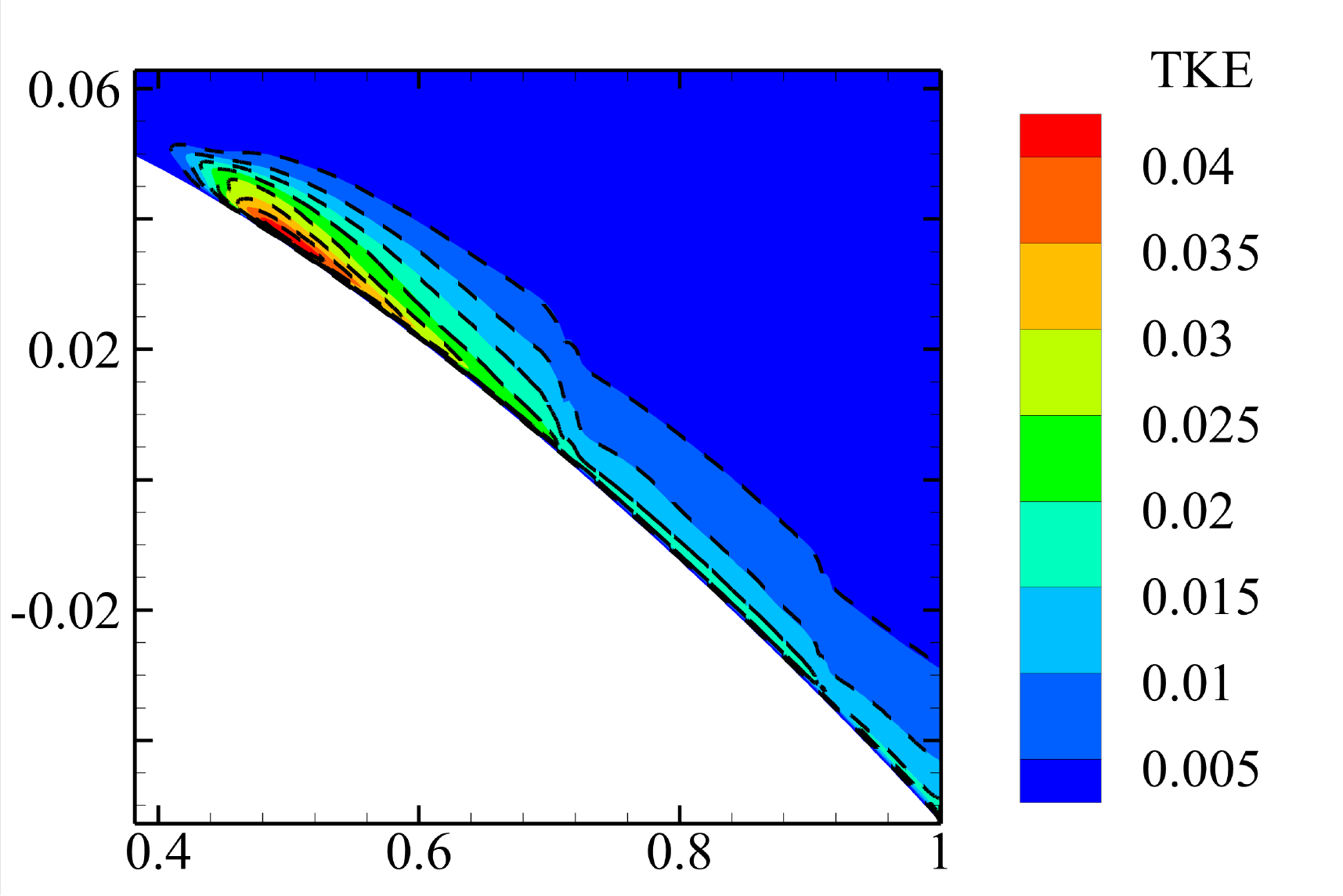}
\includegraphics[width=8cm]{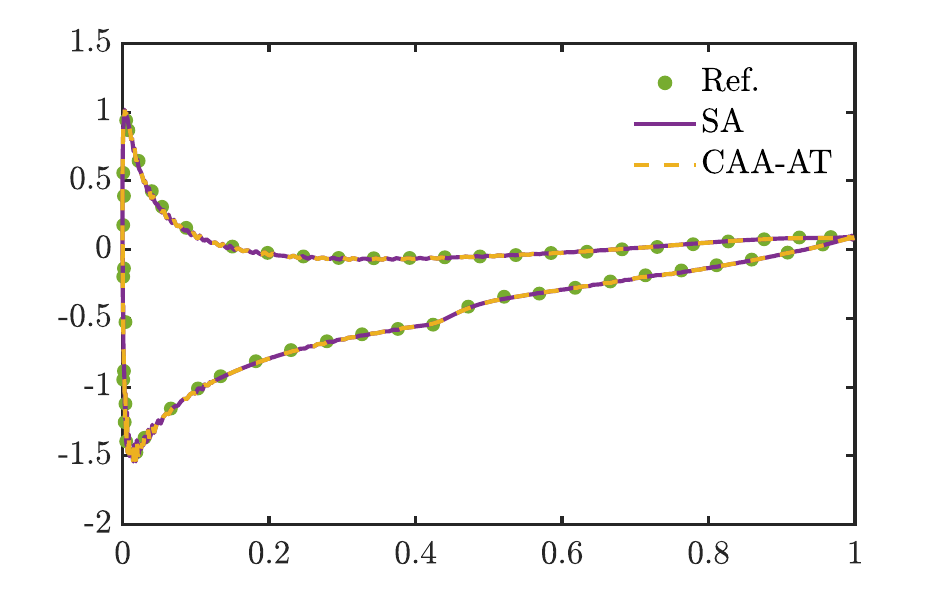}
\includegraphics[width=8cm]{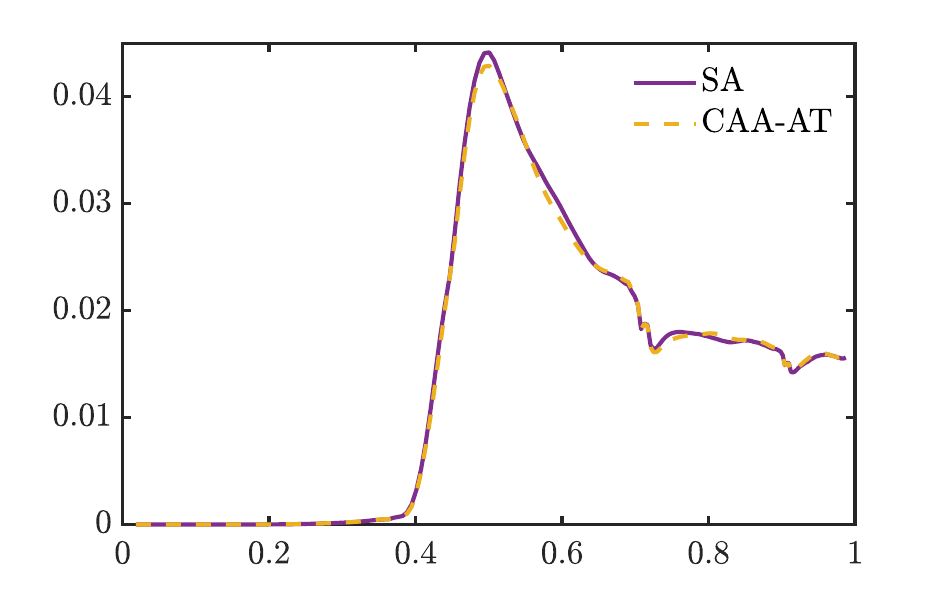}
\begin{picture}(0,0)
\put(-466,260){\bf \small (a)}
\put(-228,260){\bf \small (b)}
\put(-370,140){$x/c$}
\put(-145,140){$x/c$}
\put(-465,210){\rotatebox{90}{$y/c$}}
\put(-230,210){\rotatebox{90}{$y/c$}}
\put(-465,105){\bf \small (c)}
\put(-230,105){\bf \small (d)}
\put(-350,-5){$x/c$}
\put(-120,-5){$x/c$}
\put(-460,70){\rotatebox{90}{$c_{p}$}}
\put(-230,60){\rotatebox{90}{$\mathrm{TKE_{max}}$}}
\put(-277,120.5){\small \cite{Visbal_AIAA_2018}}
\put(-348,47){\color{black}{\vector(-1,4){4.5}}}
\put(-130,106){\color{black}{\vector(2,3){13}}}
\put(-350,40){\color{black}{Transition}}
\put(-170,100){\color{black}{Transition}}
\end{picture}
\caption{Contours of (a) pressure coefficient ($c_{p}$) and (b) turbulent kinetic energy (TKE, magnified near the suction surface) obtained from time and span averaged fields. The colored contours in the background and the overlaid dashed lines correspond to SA and CAA-AT, respectively. (c) $c_{p}$ distribution on the airfoil surface, compared against \cite{Visbal_AIAA_2018}, and (d) streamwise profile of maximum TKE along wall-normal direction ($\mathrm{TKE_{max}}$).}
\label{fig:airfoil_ave}
\end{figure}

In addition to the canonical test cases discussed in the previous sections, we now consider a practically relevant case: flow around a NACA0012 airfoil oriented at an angle of attack. A chord Reynolds number $Re_{c}=cU_{\infty}/\nu_{\infty}=2\times 10^{5}$ is considered, where $c$, $U_{\infty}$ and $\nu_{\infty}$ are the chord length, freestream velocity and viscosity at freestream conditions. Mach number $M=0.1$ and angle of attack $4\degree$ are considered to validate with the results from the high-fidelity simulations carried out in \cite{Visbal_AIAA_2018}. The 3D computational domain is represented by a C-grid containing approximately 17 million grid points, and is divided into 60 PEs along the streamwise and wall-normal directions, imposing periodicity in the spanwise direction. Inflow and outflow conditions are imposed at the far-field boundaries and a no-slip adiabatic boundary condition is imposed on the wall. Numerical simulations are carried out using SA and CAA-AT (with $L=1$) corresponding to cases 4 and 6, respectively, in Table~\ref{tab:lserk_at_schemes}. Figures~\ref{fig:airfoil_q_iso}(a) and~\ref{fig:airfoil_q_iso}(b) respectively show the instantaneous iso-surfaces of Q ($=10$) for SA and CAA-AT, colored by the $u$-velocity. Some of the PE boundaries are highlighted as translucent planes normal to the airfoil surface. The boundary layer on the airfoil suction surface is laminar near the leading edge and separates at $x/c \approx 0.3$, resulting in the formation of full-span Kelvin-Helmholtz rollers. Transition to turbulence is driven via the secondary instabilities developing over these KH rollers, resulting in small coherent structures and hairpin vortices. Notice that the iso-surfaces corresponding to CAA-AT are qualitatively identical to those of SA, indicating that the asynchronous solver consistently captures the separation-induced transition over the suction surface of the airfoil. Figure ~\ref{fig:airfoil_ave}(a,b) further compares the contours of the pressure coefficient and the turbulent kinetic energy obtained from the flow field averaged in time and span. Here, colored contours and lines represent the predictions from SA and CAA-AT, respectively. The pressure coefficient ($c_p$) on the airfoil surface and the streamwise variation of the maximum turbulent kinetic energy ($\mathrm{TKE_{max}}$) are compared in Fig ~\ref{fig:airfoil_ave}(c,d). CAA-AT is in excellent agreement with SA in capturing the pressure distribution, especially near the transition region, whereas good agreement is observed in the case of TKE. It should be noted that both schemes predict a consistent transition location and $\mathrm{TKE_{max}}$. This demonstrates the robustness of the CAA-AT schemes in predicting sensitive phenomena like transition, while achieving higher scalability than SA.

\begin{figure}[]
\centering
\includegraphics[width=6cm]{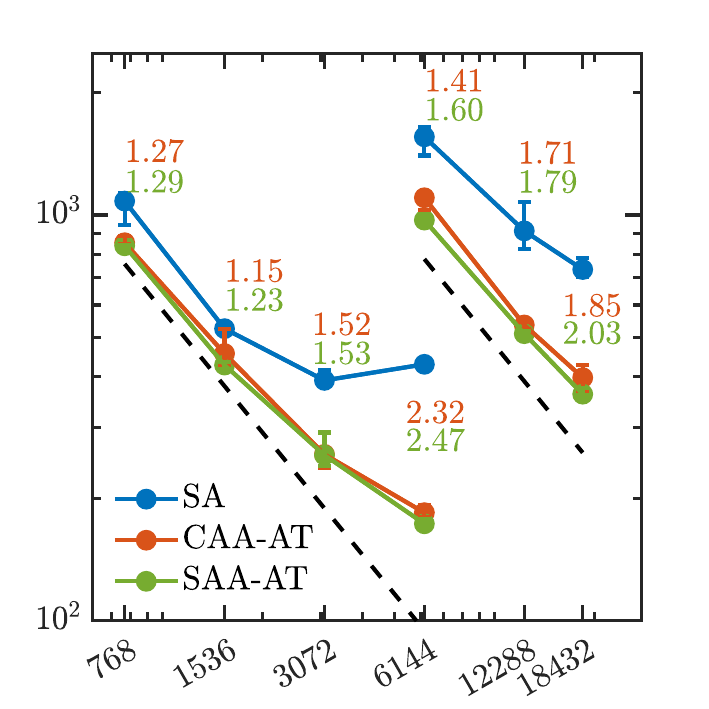}\hspace{1cm}
\includegraphics[width=6cm]{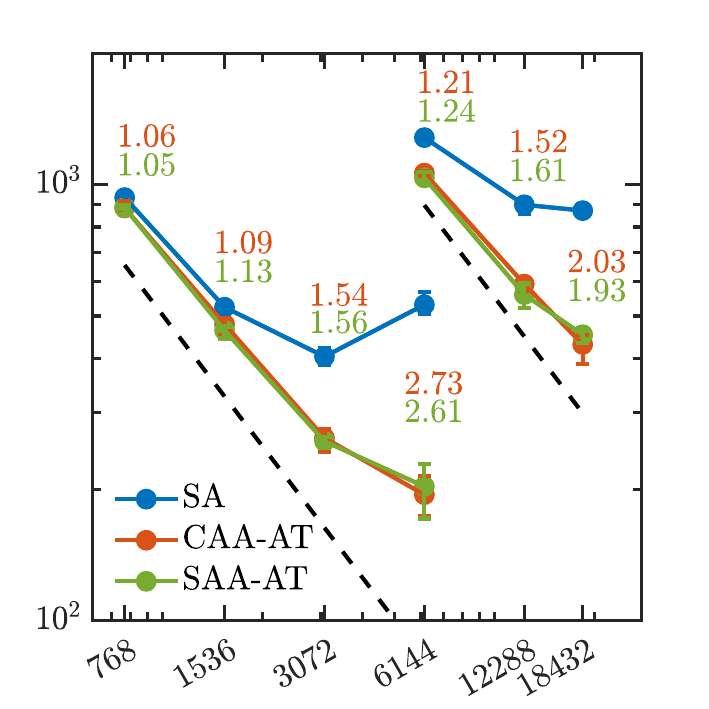}
\begin{picture}(0,0)
    \put(-385,150){\bf \small (a)}
    \put(-180,150){\bf \small (b)}
    \put(-325,-5){\small Number of PEs}
    \put(-120,-5){\small Number of PEs}
    \put(-385,40){\rotatebox{90}{\small Total execution time (s)}}
    \put(-180,40){\rotatebox{90}{\small Total execution time (s)}}
\end{picture}
\caption{Strong scaling graphs for two resolutions of the airfoil case on (a) PARAM Pravega and (b) PARAM Rudra supercomputers, for SA, CAA-AT and SAA-AT (see cases 4 and 6 in Table~\ref{tab:lserk_at_schemes}). Maximum allowable delay $L=5$ for CAA-AT and SAA-AT.}
\label{fig:airfoil_scaling}
\end{figure}

\begin{figure}[]
\centering
\includegraphics[width=6cm]{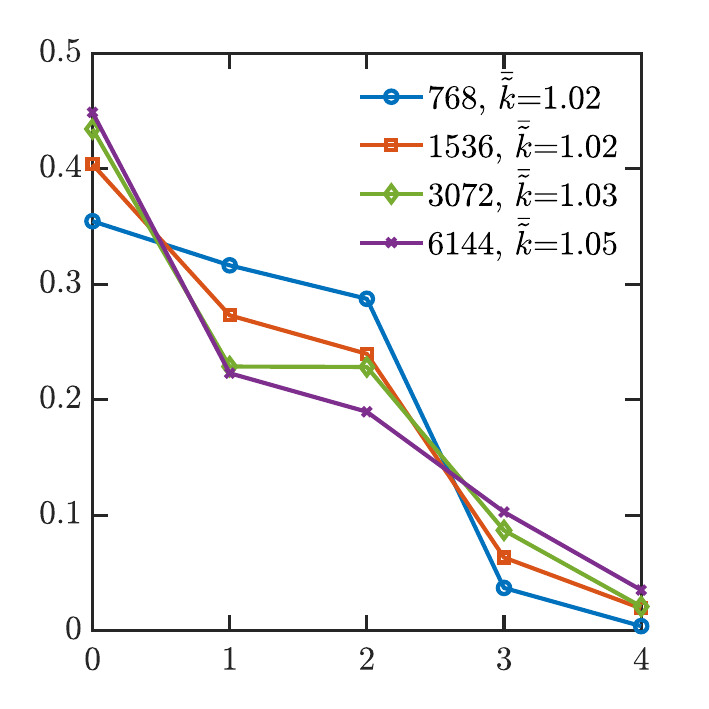}\hspace{1cm}
\includegraphics[width=6cm]{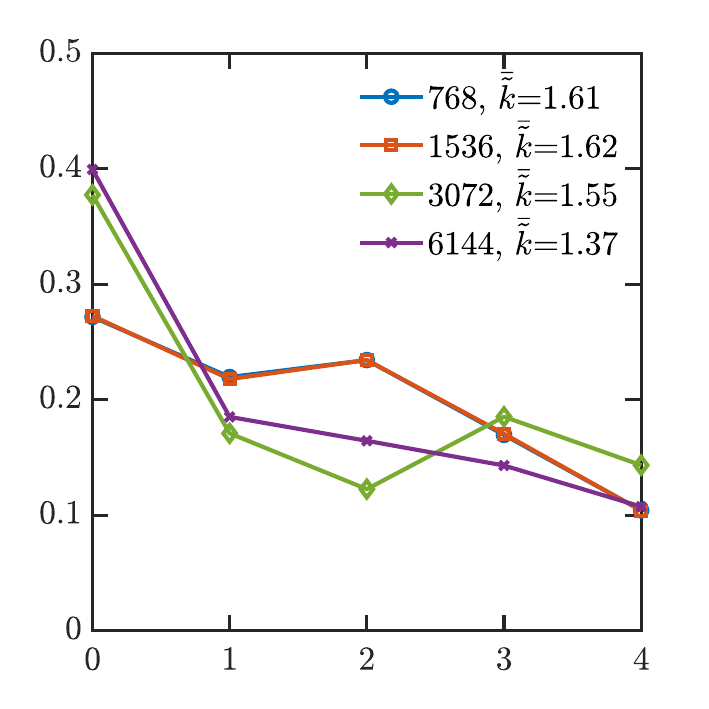}
\begin{picture}(0,0)
    \put(-385,150){\bf \small (a)}
    \put(-180,150){\bf \small (b)}
    \put(-305,-5){\small Delay $\tilde{k}$}
    \put(-100,-5){\small Delay $\tilde{k}$}
    \put(-385,60){\rotatebox{90}{\small Probability}}
    \put(-180,60){\rotatebox{90}{\small Probability}}
\end{picture}
\caption{Probability distribution of delays for the airfoil case on (a) PARAM Pravega and (b) PARAM Rudra supercomputers, for SAA-AT (see case 6 in Table~\ref{tab:lserk_at_schemes}). Maximum allowable delay $L=5$ for SAA-AT.}
\label{fig:airfoil_saa_delay}
\end{figure}

From a computational performance standpoint, the domain decomposition of complex geometries, like the airfoil, leads to nonuniform workload distribution among PEs. These workload imbalances incur additional overheads and lead to performance variation among PEs. The performance of the asynchronous solver needs to be assessed in such practical conditions. Strong scaling experiments with two resolutions, containing 2.08 billion and 18 billion grid points with two-dimensional decomposition, are performed on the PARAM Pravega and PARAM Rudra supercomputers. The benchmarks are performed on up to 18,432 PEs by evolving for 100 time steps over 10 independent trials to quantify the performance variation, considering a maximum allowable delay of 5 for the asynchronous solvers. As predicted, the strong scaling results of both CAA-AT and SAA-AT lie closer to the ideal scaling line compared to SA. The asynchronous algorithms report speed-ups of $2.47\times$ and $2.73\times$ on PARAM Pravega (Fig.~\ref{fig:airfoil_scaling}(a)) and PARAM Rudra (Fig.~\ref{fig:airfoil_scaling}(b)), respectively, relative to SA. For SAA-AT, the behavior of communication delays plays a major role in reducing the synchronization overheads among PEs. Here, we quantify the behavior of delays using a probability distribution across different PE counts on both supercomputers as shown by the plots in Fig.~\ref{fig:airfoil_saa_delay}. The overall trend is that smaller values of delay are more likely than larger delays, with the mean delay lying between 1 and 2 time levels. At higher PE counts, data movement increases in the network, thereby increasing the probability of incurring higher delays. An interesting comparison can be made between the two architectures on the correlation between the delay distributions and the scaling graphs. At the extreme scale (6144 PEs), we observe higher probabilities for larger delays in the case of PARAM Rudra (Fig.~\ref{fig:airfoil_saa_delay}(b)) compared to PARAM Pravega (Fig.~\ref{fig:airfoil_saa_delay}(a)), leading to a greater deviation in the execution time for the former (Fig.~\ref{fig:airfoil_scaling}(b)) compared to the latter (Fig.~\ref{fig:airfoil_scaling}(a)). This demonstrates the dependence of communication delays on factors like workload distribution and the architecture.

\section{Conclusions\label{sec:conclusions}}
The asynchronous computing approach has recently been explored to address the communication overheads in parallel PDE solvers. These communication overheads are considered to be the major bottleneck to scalability in state-of-the-art DNS solvers. This provides the motivation to assess the efficacy of the asynchronous computing approach in accurately capturing flow properties while reducing the communication overheads in CFD solvers. In this study, high-order asynchrony-tolerant (AT) schemes that relax communication/synchronization at a mathematical level using delayed data at PE boundaries were implemented into the multi-block structured compressible flow solver COMP-SQUARE using two algorithms that introduce asynchrony in a distributed parallel setting. These algorithms also skip communications at intermediate stages of time advancement when Runge-Kutta schemes are used. The communication avoiding algorithm (CAA) relaxes communication by performing communications periodically once every few time steps, characterized by a uniform periodic delay across all PEs. The synchronization avoiding algorithm (SAA), on the other hand, relaxes the synchronization by initiating communication at every time step without enforcing explicit synchronization. Numerical experiments were performed using three test cases to evaluate the efficacy of the algorithms. First, the canonical problem of isentropic vortex advection was solved to demonstrate the behavior of numerical errors due to communication delays and the overall order of accuracy of the AT schemes. The computational performance of the asynchronous algorithms was compared against the baseline synchronous algorithm (SA) using a strong scaling analysis, where the considerable speed-ups of CAA-AT and SAA-AT relative to SA are demonstrated. This is a direct consequence of relaxed communication/synchronization. The second test case is the three-dimensional Taylor-Green vortex problem, which is a decaying vortex field transitioning into a turbulent flow. Important flow features were compared between SA, CAA-AS, and CAA-AT, showing the emergence of spurious numerical oscillations for CAA-AS (standard scheme used with delayed data) and excellent agreement between CAA-AT and SA. The computational performance was assessed by performing strong and weak scaling studies, and speed-ups of up to $4\times$ were achieved for the asynchronous solver relative to the SA-based solver. Thirdly, a more practical case of flow around a NACA0012 airfoil was considered. The asynchronous solver accurately captures the separation induced transition as well as the small scale structures in the turbulent boundary layer, suggesting that the high frequency numerical errors due to communication delays are reduced using the higher-order AT schemes near PE boundaries. Further, average flow features such as pressure and turbulent kinetic energy distribution were compared. The pressure distribution obtained with CAA-AT was very similar to that of SA, while a good agreement was observed in turbulent kinetic energy. The computational performance was assessed using strong scaling experiments carried out on two different architectures, achieving up to $2.73\times$ speed-up. Furthermore, the probability distribution of delays was also obtained to understand the stochastic behavior of delays in SAA-AT. The findings of this study demonstrate the applicability of asynchronous computing approach in contemporary scalable CFD codes to accelerate numerical simulations that can leverage the computing abilities of future exascale machines.

\section*{Acknowledgements}
The authors thank C.P. Abdul Gafoor for his help in setting up the airfoil test case. KA acknowledges the financial support from the MoE-STARS grant and the ANRF Core Research Grant. The authors thank the National Supercomputing Mission, India, P.G. Senapathy Centre, IIT Madras and SERC, IISc for providing the computing resources. The authors thank the National PARAM Supercomputing Facility (NPSF) for providing the computing resources on the PARAM Rudra cluster under the National Supercomputing Mission.

\section*{Data Availability Statement}

The data that support the findings of this study are available from the corresponding author upon reasonable request.

\appendix

\section{Flux terms in governing equations}
\label{sec:flux_math}

The unsteady conservation of mass, momentum, and energy equations are non-dimensionalized and cast in strong conservative form on general curvilinear coordinates $(x,y,z)\rightarrow(\xi,\eta,\zeta)$, as detailed in \cite{Gaitonde_1998,Visbal_JCP_2002}.
\begin{equation}
    \frac{\partial}{\partial t}\left(\frac{\bm{U}}{J}\right) + \frac{\partial \bm{\hat{F}}}{\partial \xi} + \frac{\partial \bm{\hat{G}}}{\partial \eta} + \frac{\partial \bm{\hat{H}}}{\partial \zeta} = \frac{1}{Re}\left[\frac{\partial \bm{\hat{F}}_{v}}{\partial \xi} + \frac{\partial \bm{\hat{G}}_{v}}{\partial \eta} + \frac{\partial \bm{\hat{H}}_{v}}{\partial \zeta}\right],
    \label{eq:NS_eqn_appn}
\end{equation}
where $\bm{U} = \{\rho,\rho u, \rho v, \rho w, \rho E_t\}$ is the vector of the conserved variables and $J = \partial (\xi, \eta, \zeta)/\partial (x, y, z)$ is the coordinate transformation Jacobian. The advective fluxes $\bm{\hat{F}}$, $\bm{\hat{G}}$, and $\bm{\hat{H}}$ are
\begin{equation}
\bm{\hat{F}} =    \begin{bmatrix}
\rho \hat{U}\\
\rho u \hat{U} + \hat{\xi}_xp\\
\rho v \hat{U} + \hat{\xi}_yp\\
\rho w \hat{U} + \hat{\xi}_zp\\
(\rho E_t + p)\hat{U}
\end{bmatrix},\
\bm{\hat{G}} =    \begin{bmatrix}
\rho \hat{V}\\
\rho u \hat{V} + \hat{\eta}_xp\\
\rho v \hat{V} + \hat{\eta}_yp\\
\rho w \hat{V} + \hat{\eta}_zp\\
(\rho E_t + p)\hat{V}
\end{bmatrix},\
\bm{\hat{H}} =    \begin{bmatrix}
\rho \hat{W}\\
\rho u \hat{W} + \hat{\zeta}_xp\\
\rho v \hat{W} + \hat{\zeta}_yp\\
\rho w \hat{W} + \hat{\zeta}_zp\\
(\rho E_t + p)\hat{W}
\end{bmatrix}.
\label{eq:inv_flux}
\end{equation}
Here, $\hat{\xi}_{x}=J^{-1}\partial\xi/\partial x$ is the metric term, $\hat{U}$, $\hat{V}$ and $\hat{W}$ are the contravariant components of velocity, $p$ is the static pressure, and $E_t$ is the total specific energy.
\begin{equation*}
\begin{split}
    \hat{U} &= \hat{\xi}_xu + \hat{\xi}_yv + \hat{\xi}_zw\\ 
    \hat{V} &= \hat{\eta}_xu + \hat{\eta}_yv + \hat{\eta}_zw\\
    \hat{W} &= \hat{\zeta}_xu + \hat{\zeta}_yv + \hat{\zeta}_zw\\
    E_t &= \frac{T}{\gamma(\gamma-1)M_{\infty}^{^2}} + \frac{1}{2}(u^2 + v^2 + w^2).
\end{split}
\label{eq:contra_vel}
\end{equation*}
The viscous fluxes $\bm{\hat{F}}_{v}$, $\bm{\hat{G}}_{v}$, and $\bm{\hat{H}}_{v}$ involve second-order derivatives and are written as
\begin{equation}
\bm{\hat{F}}_{v} =  \frac{1}{J}  \begin{bmatrix}
0 \\
\xi_i \tau_{i1}\\
\xi_i \tau_{i2}\\
\xi_i \tau_{i3}\\
\xi_i b_{i}
\end{bmatrix},\
\bm{\hat{G}}_{v} =  \frac{1}{J}  \begin{bmatrix}
0\\
\eta_i \tau_{i1}\\
\eta_i \tau_{i2}\\
\eta_i \tau_{i3}\\
\eta_i b_{i}
\end{bmatrix},\
\bm{\hat{H}}_{v} =   \frac{1}{J}
\begin{bmatrix}
0\\
\zeta_i \tau_{i1}\\
\zeta_i \tau_{i2}\\
\zeta_i \tau_{i3}\\
\zeta_i b_{i}
\end{bmatrix},
\label{eq:visc_flux}
\end{equation}
where the stress tensor $\tau_{ij}$ and heat flux vector $b_i$ are
\begin{equation}
\begin{aligned}
    \tau_{ij} &= \mu\left(\frac{\partial \xi_k}{\partial x_j}\frac{\partial u_i}{\partial \xi_k} + \frac{\partial \xi_k}{\partial x_i}\frac{\partial u_j}{\partial \xi_k}\right) - \frac{2}{3}\mu\delta_{ij}\frac{\partial \xi_l}{\partial x_i}\frac{\partial u_k}{\partial \xi_l},\\
    b_i &= u_{j}\tau_{ij} + \frac{\mu}{(\gamma-1)Re Pr M_{\infty}^2}\frac{\partial \xi_l}{\partial x_i}\frac{\partial T}{\partial \xi_l}.
\end{aligned}
\label{eq:viscous_gradients}
\end{equation}

\section{Standard finite-difference and Runge-Kutta schemes used in COMP-SQUARE}
\label{app:fd-schemes}
Let us recall the general expression to obtain the spatial derivative of a function $\phi$,
\begin{equation*}
\alpha \phi^{'}_{i-1} + \phi^{'}_{i} + \alpha \phi^{'}_{i+1} = b\frac{\phi_{i+2}-\phi_{i-2}}{4\Delta\xi} + a\frac{\phi_{i+1}-\phi_{i-1}}{2\Delta\xi}.
\label{eq:spat_disc_appn} 
\end{equation*}
The parameters $\alpha$, $a$, and $b$ for the spatial schemes used in this paper are as follows.
\begin{itemize}
    \item \textbf{CD2:} $\alpha=0$, $a=1$, $b=0$
    \item \textbf{CD4:} $\alpha=0$, $a=4/3$, $b=-1/3$
\end{itemize}
The low-storage explicit Runge-Kutta (LSERK) update equations are of the form
\begin{equation*}
\begin{split}
\bm{Q}^{(m)} &= A_m\bm{Q}^{(m-1)} + \Delta t \bm{R}(\bm{U}^{(m-1)})\\     \bm{U}^{(m)} &= \bm{U}^{(m-1)} + B_m \bm{Q}^{(m)},
\end{split}
\label{eq:lserk_Q_appn}
\end{equation*}
where $\bm{U}$ is the vector of conserved variables and $\bm{Q}$ is the intermediate vector. The coefficients $A_m$ and $B_m$ for the LSERK2 and LSERK4 schemes, respectively, are $A_m \in \{0,-1/2\},\ B_m \in \{1/2,1\}$ and as given in Table~\ref{tab:lserk4_coeffs}.
\begin{table}[h!]
\centering
\begin{tabular}{c c c}
\hline
$m$ & $A_m$ & $B_m$ \\
\hline
$1$ & $0$ & $0.09761835$ \\
$2$ & $-0.4812317$ & $0.41225329$ \\
$3$ & $-1.0495626$ & $0.44021696$ \\
$4$ & $-1.6025296$ & $1.42631146$ \\
$5$ & $-1.7782672$ & $0.19787605$ \\
\hline
\end{tabular}
\caption{Coefficients for the five-stage LSERK4 method, adopted from \cite{Carpenter_report_1994}.}
\label{tab:lserk4_coeffs}
\end{table}

\section{Asynchrony-tolerant (AT) update schemes for buffer points\label{sec:at_buffer_appendix}}

For the LSERK2-CD2-AT2 method, the solution at a buffer point must be updated at stage $m=1$ alone, which requires the computation of $\partial \phi/\partial t\big|_{i}^{(0)}$ based on the extended stencil in time. Solving Eq.~\ref{eq:at_buffer_constraints} in Sec.~\ref{subsec:at_schemes} for the unknown coefficients results in the approximation
\begin{equation*}
(\partial \phi/\partial t)\Biggr|_{i}^{(0)} = \dfrac{\left(2\tilde{k}+3\right)}{2\Delta t}\phi_{i}^{n-\tilde{k}} - \dfrac{\left(2\tilde{k}+2\right)}{\Delta t}\phi_{i}^{n-\tilde{k}-1} + \dfrac{\left(2\tilde{k}+1\right)}{2\Delta t}\phi_{i}^{n-\tilde{k}-2}.
\end{equation*}
On the other hand, for the LSERK4-CD4-AT4 method, buffer updates should be performed until the fourth stage ($m=4$) and the expressions obtained for the temporal derivative in Eq.~\ref{eq:lserk_at_buffer} are listed in Table~\ref{tab:buff_update_coeffs}.

\begin{table}[h!]
    \centering
    \begin{tabular}{l l}
    \hline
       Derivative  & Expression  \\
       \hline
       \multirow{5}{*}{$(\partial \phi/\partial t)\Big|_{i}^{(0)}$}
        &  $\dfrac{1}{12\Delta t}\left(2\tilde{k}^3+15\tilde{k}^2+35\tilde{k}+25\right)\phi_{i}^{n-\tilde{k}}$\\[1.2ex]
        &  $-\dfrac{1}{6\Delta t}\left(4\tilde{k}^3+27\tilde{k}^2+52\tilde{k}+24\right)\phi_{i}^{n-\tilde{k}-1}$ \\[1.2ex]
        &  $+\dfrac{1}{2\Delta t}\left(2\tilde{k}^3+12\tilde{k}^2+19\tilde{k}+6\right)\phi_{i}^{n-\tilde{k}-2}$ \\[1.2ex]
        &  $-\dfrac{1}{6\Delta t}\left(4\tilde{k}^3+21\tilde{k}^2+28\tilde{k}+8\right)\phi_{i}^{n-\tilde{k}-3}$ \\[1.2ex]
        &  $+\dfrac{1}{12\Delta t}\left(2\tilde{k}^3+9\tilde{k}^2+11\tilde{k}+3\right)\phi_{i}^{n-\tilde{k}-4}$ \\\\
        
        \multirow{5}{*}{$(\partial \phi/\partial t)\Big|_{i}^{(1)}$}
        &  $\dfrac{1}{12\Delta t}\left(2\tilde{k}^3+(15+6\nu_{1}^{1})\tilde{k}^2+(35+30\nu_{1}^{1})\tilde{k}+25+35\nu_{1}^{1}\right)\phi_{i}^{n-\tilde{k}}$\\[1.2ex]
        &  $-\dfrac{1}{6\Delta t}\left(4\tilde{k}^3+(27+12\nu_{1}^{1})\tilde{k}^2+(52+54\nu_{1}^{1})\tilde{k}+24+52\nu_{1}^{1}\right)\phi_{i}^{n-\tilde{k}-1}$ \\[1.2ex]
        &  $+\dfrac{1}{2\Delta t}\left(2\tilde{k}^3+(12+6\nu_{1}^{1})\tilde{k}^2+(19+24\nu_{1}^{1})\tilde{k}+6+19\nu_{1}^{1}\right)\phi_{i}^{n-\tilde{k}-2}$ \\[1.2ex]
        &  $-\dfrac{1}{6\Delta t}\left(4\tilde{k}^3+(21+12\nu_{1}^{1})\tilde{k}^2+(28+42\nu_{1}^{1})\tilde{k}+8+28\nu_{1}^{1}\right)\phi_{i}^{n-\tilde{k}-3}$ \\[1.2ex]
        &  $+\dfrac{1}{12\Delta t}\left(2\tilde{k}^3+(9+6\nu_{1}^{1})\tilde{k}^2+(11+18\nu_{1}^{1})\tilde{k}+3+11\nu_{1}^{1}\right)\phi_{i}^{n-\tilde{k}-4}$ \\\\

        \multirow{5}{*}{$(\partial \phi/\partial t)\Big|_{i}^{(2)}$}
        &  $\dfrac{1}{12\Delta t}\left(2\tilde{k}^3+(15+6\nu_{1}^{2})\tilde{k}^2+(35+30\nu_{1}^{2}+12\nu_{2}^{2})\tilde{k}+25+35\nu_{1}^{2}+30\nu_{2}^{2}\right)\phi_{i}^{n-\tilde{k}}$\\[1.2ex]
        &  $-\dfrac{1}{6\Delta t}\left(4\tilde{k}^3+(27+12\nu_{1}^{2})\tilde{k}^2+(52+54\nu_{1}^{2}+24\nu_{2}^{2})\tilde{k}+24+52\nu_{1}^{2}+54\nu_{2}^{2}\right)\phi_{i}^{n-\tilde{k}-1}$ \\[1.2ex]
        &  $+\dfrac{1}{2\Delta t}\left(2\tilde{k}^3+(12+6\nu_{1}^{2})\tilde{k}^2+(19+24\nu_{1}^{2}+12\nu_{2}^{2})\tilde{k}+6+19\nu_{1}^{2}+24\nu_{2}^{2}\right)\phi_{i}^{n-\tilde{k}-2}$ \\[1.2ex]
        &  $-\dfrac{1}{6\Delta t}\left(4\tilde{k}^3+(21+12\nu_{1}^{2})\tilde{k}^2+(28+42\nu_{1}^{2}+24\nu_{2}^{2})\tilde{k}+8+28\nu_{1}^{2}+42\nu_{2}^{2}\right)\phi_{i}^{n-\tilde{k}-3}$ \\[1.2ex]
        &  $+\dfrac{1}{12\Delta t}\left(2\tilde{k}^3+(9+6\nu_{1}^{2})\tilde{k}^2+(11+18\nu_{1}^{2}+12\nu_{2}^{2})\tilde{k}+3+11\nu_{1}^{2}+18\nu_{2}^{2}\right)\phi_{i}^{n-\tilde{k}-4}$ \\\\

        \multirow{5}{*}{$(\partial \phi/\partial t)\Big|_{i}^{(3)}$}
        &  $\dfrac{1}{12\Delta t}\left(2\tilde{k}^3+(15+6\nu_{1}^{3})\tilde{k}^2+(35+30\nu_{1}^{3}+12\nu_{2}^{3})\tilde{k}+25+35\nu_{1}^{3}+30\nu_{2}^{3}+12\nu_{3}^{3}\right)\phi_{i}^{n-\tilde{k}}$\\[1.2ex]
        &  $-\dfrac{1}{6\Delta t}\left(4\tilde{k}^3+(27+12\nu_{1}^{3})\tilde{k}^2+(52+54\nu_{1}^{3}+24\nu_{2}^{3})\tilde{k}+24+52\nu_{1}^{3}+54\nu_{2}^{3}+24\nu_{3}^{3}\right)\phi_{i}^{n-\tilde{k}-1}$ \\[1.2ex]
        &  $+\dfrac{1}{2\Delta t}\left(2\tilde{k}^3+(12+6\nu_{1}^{3})\tilde{k}^2+(19+24\nu_{1}^{3}+12\nu_{2}^{3})\tilde{k}+6+19\nu_{1}^{3}+24\nu_{2}^{3}+12\nu_{3}^{3}\right)\phi_{i}^{n-\tilde{k}-2}$ \\[1.2ex]
        &  $-\dfrac{1}{6\Delta t}\left(4\tilde{k}^3+(21+12\nu_{1}^{3})\tilde{k}^2+(28+42\nu_{1}^{3}+24\nu_{2}^{3})\tilde{k}+8+28\nu_{1}^{3}+42\nu_{2}^{3}+24\nu_{3}^{3}\right)\phi_{i}^{n-\tilde{k}-3}$ \\[1.2ex]
        &  $+\dfrac{1}{12\Delta t}\left(2\tilde{k}^3+(9+6\nu_{1}^{3})\tilde{k}^2+(11+18\nu_{1}^{3}+12\nu_{2}^{3})\tilde{k}+3+11\nu_{1}^{3}+18\nu_{2}^{3}+12\nu_{3}^{3}\right)\phi_{i}^{n-\tilde{k}-4}$ \\\\
        \hline
    \end{tabular}
    \caption{Temporal derivatives of function $\phi$ at buffer points based on the delay $\tilde{k}$, evaluated at $(m-1)$th stage in the LSERK4-CD4-AT4.}
    \label{tab:buff_update_coeffs}
\end{table}

\bibliographystyle{main-style}
\bibliography{main}

\end{document}